\documentclass[epj]{svjour}
\usepackage{graphics,multirow}
\usepackage{ulem}
\begin{document}
\title{Study of the reaction \boldmath ${\gamma p \rightarrow K^+ \Lambda(1520)}$ \unboldmath
at photon energies up to 2.65\,GeV\thanks{This work is supported in
part by the Deutsche Forschungsgemeinschaft (SPP KL 980/2-3) and the
SFB/TR16}}
\titlerunning{Measurement of the reaction $\gamma p \rightarrow K^ + \Lambda(1520)$
at photon energies up to $2.65$\,GeV}
\author{
F.\,W.\,Wieland\inst{1,4},
J.\,Barth\inst{1},
K.-H.\,Glander\inst{1,6},
J.\,Hannappel\inst{1},
N.\,J\"open\inst{1},
F.\,Klein\inst{1,}\thanks{E-mail: klein@physik.uni-bonn.de},
E.\,Klempt\inst{2},
R.\,Lawall\inst{1,7},
D.\,Menze\inst{1},
M.\,Ostrick\inst{3}
E.\,Paul\inst{1},
I.\,Schulday\inst{1,5},
W.\,J.\,Schwille\inst{1}
}
\authorrunning{F.\,W.\,Wieland et al.}
\institute{
Physikalisches Institut der Universit\"at Bonn, Germany
\and Helmholtz-Institut f\"ur Strahlen- und Kernphysik,  Universit\"at Bonn, Germany
\and Institut f\"{ur} Kernphysik, Universit\"at Mainz, Germany
\and presently Wieland KG, L\"{u}denscheid, Germany
\and presently IFB AG, K\"{o}ln, Germany
\and presently TRW Automotive GmbH, Alfdorf, Germany
\and presently T\"{U}V Nord, Germany
}
\date{Received: date / Revised version: date}
\abstract{ The reaction  $\gamma p \rightarrow K^+\Lambda(1520)$ was
measured in the energy range from threshold to 2.65\,GeV with the
SAPHIR detector at the electron stretcher facility ELSA in Bonn. The
$\Lambda(1520)$ production cross section was analyzed in the decay
modes $pK^-$, $n \bar{K}^0$, $\Sigma^{\pm}\pi^{\mp}$, and
$\Lambda\pi^+\pi^-$ as a function of the photon energy and the
squared four-momentum transfer $t$. While the cross sections for the
inclusive reactions rise steadily with energy, the cross section of
the process $\gamma p \rightarrow K^+\Lambda(1520)$ peaks at a
photon energy of about $2.0$\,GeV, falls off exponentially with $t$,
and shows a slope flattening with increasing photon energy. The
angular distributions in the $t$--channel helicity system indicate
neither a $K$ nor a $K^\star$ exchange dominance. The interpretation
of the $\Lambda(1520)$ as a $\Sigma(1385)\pi$ molecule is not
supported.
 % end of PACS codes
} %end of abstract
\maketitle
\parindent0mm
%\begin{sloppypar}

\section{Introduction}
\label{intro}

We report on a study of the dynamics of the process $\gamma p
\rightarrow K^+ \Lambda(1520)$ for photon energies from threshold to
2.65\,GeV. The CLAS collaboration \cite{Barrow:2001ds} investigated
this process in electroproduction, at electron beam energies of
4.05, 4.25, and 4.46\,GeV, in the kinematic region spanning the
squared momentum transfer $Q^2$ from $0.9$ to $2.4$\,GeV$^2$ and for
invariant masses $W$ from $1.95$ to $2.65$\,GeV, and suggested that
the reaction is dominated by $t$-channel processes. From a
comparison between their $t$-channel helicity-frame angular
distributions and angular distributions from photoproduction
measured at Daresbury for photon energies from 2.8 to 4.8\,GeV
\cite{daresbury}, the CLAS collaboration concluded that $t$-channel
diagrams with longitudinally polarized photons contribute
significantly to electroproduction of $\Lambda(1520)$. Here we
present new photoproduction results in a $W$ range below the
Daresbury data, allowing for a better comparison with CLAS.

The $\Lambda(1520)$ production was investigated in the decay
channels $\Lambda(1520)\rightarrow pK^-$, $nK^0_s$,
$\Sigma^\pm\pi^\mp$, and $\Lambda\pi^+\pi^-$. We present, in four
energy bins, total cross sections, differential cross sections
$d\sigma/dt$ and angular distributions in the $t$-channel helicity
system. $\Lambda$ decay fractions and an upper limit for the
fractional contribution of $\Lambda(1520)\rightarrow\Sigma(1385)\pi$
in the $\Lambda\pi^+\pi^-$ final state are reported. The results
presented here are based on 180 million triggers which were taken
with the SAPHIR detector. A study of the related reactions $\gamma p
\rightarrow K^+ \Sigma^{\pm}\pi^{\mp}$ can be found in an
accompanying paper \cite{Schulday:}. The data are available via
internet.\footnote{http://saphir.physik.uni-bonn.de/saphir/publications}

\section{The experiment}
\label{saphir}

\begin{figure*}[ht!]
\hspace{4cm}
\resizebox{0.6\textwidth}{!}{%
  \rotatebox{270}{\includegraphics{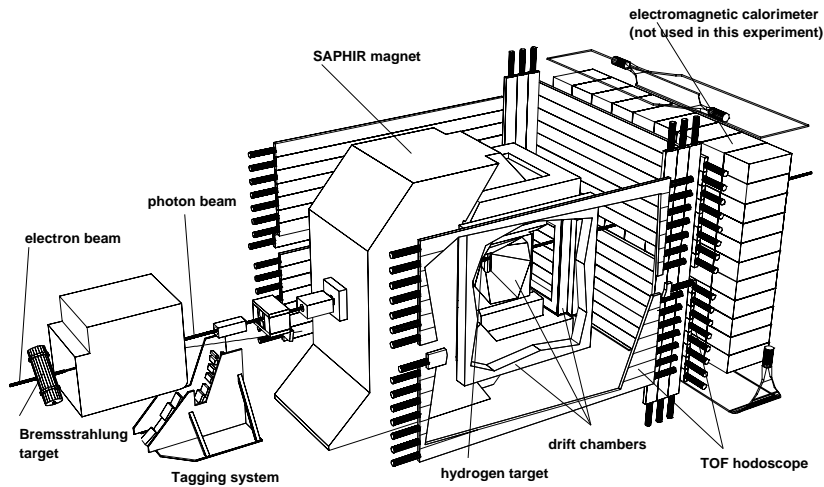}}
}
\caption{Sketch of the SAPHIR detector.}
\label{pic:saphir}
\end{figure*}

The SAPHIR detector is a magnetic multiparticle spectrometer
\cite{Schwille} which took data at the 3.5\,GeV electron stret\-cher
facility ELSA \cite{Husmann}. The setup is shown schematically in
fig. \ref{pic:saphir}. An extracted electron beam of 2.8\,GeV was
directed on a radiator target to provide an energy-tagged photon
beam within the range from 0.868\,GeV to 2.650\,GeV. The data-taking
was based on a trigger defined by a coincidence of signals from the
scattered electrons in the tagging system with at least two charged
particles in the scintillator hodoscopes, and no signal from a beam
veto counter downstream of SAPHIR which detected non-interacting
photons. The drift chambers served to measure charged particle
tracks. The scintillator hodoscopes were used for time-of-flight
(TOF) measurements and in order to determine the particle masses
from the track momentum (measured with the drift chambers in the
magnetic field). For a more detailed description see
\cite{Barth,Glander03b}.

\section{Event reconstruction and event selection}
\label{selection}

\subsection{\boldmath $\Lambda(1520) \rightarrow pK^-$ \unboldmath}

Events from the reaction $\gamma p \rightarrow K^+pK^-$ were
selected from three-track events via a kinematic fit using the
measured photon energy in the tagging system and the momenta of the
charged particles.  Events were accepted, if the fit probability
$P(\chi^2)$ was larger than those of competing reactions (see table
\ref{tab:reactions}) and  larger than 0.001, and if the mass
assignments of the charged  particles were in accordance with the
time-of-flight measurement in the scintillator hodoscopes. In order
to achieve a better signal to background ratio, events from the
reaction $\gamma p \rightarrow p\phi$ with $\phi$ decays into
$K^+K^-$ were removed for photon energies above 1.93\,GeV by a cut
in the $K^+K^-$ mass distribution ($1.0097$\,GeV
$<M_{K^+K^-}<1.0285$\,GeV).

\begin{figure*}[ht!] \hspace{0cm}
\begin{tabular}{c c}
\resizebox{0.5\textwidth}{!}{%
  \rotatebox{0}{\includegraphics{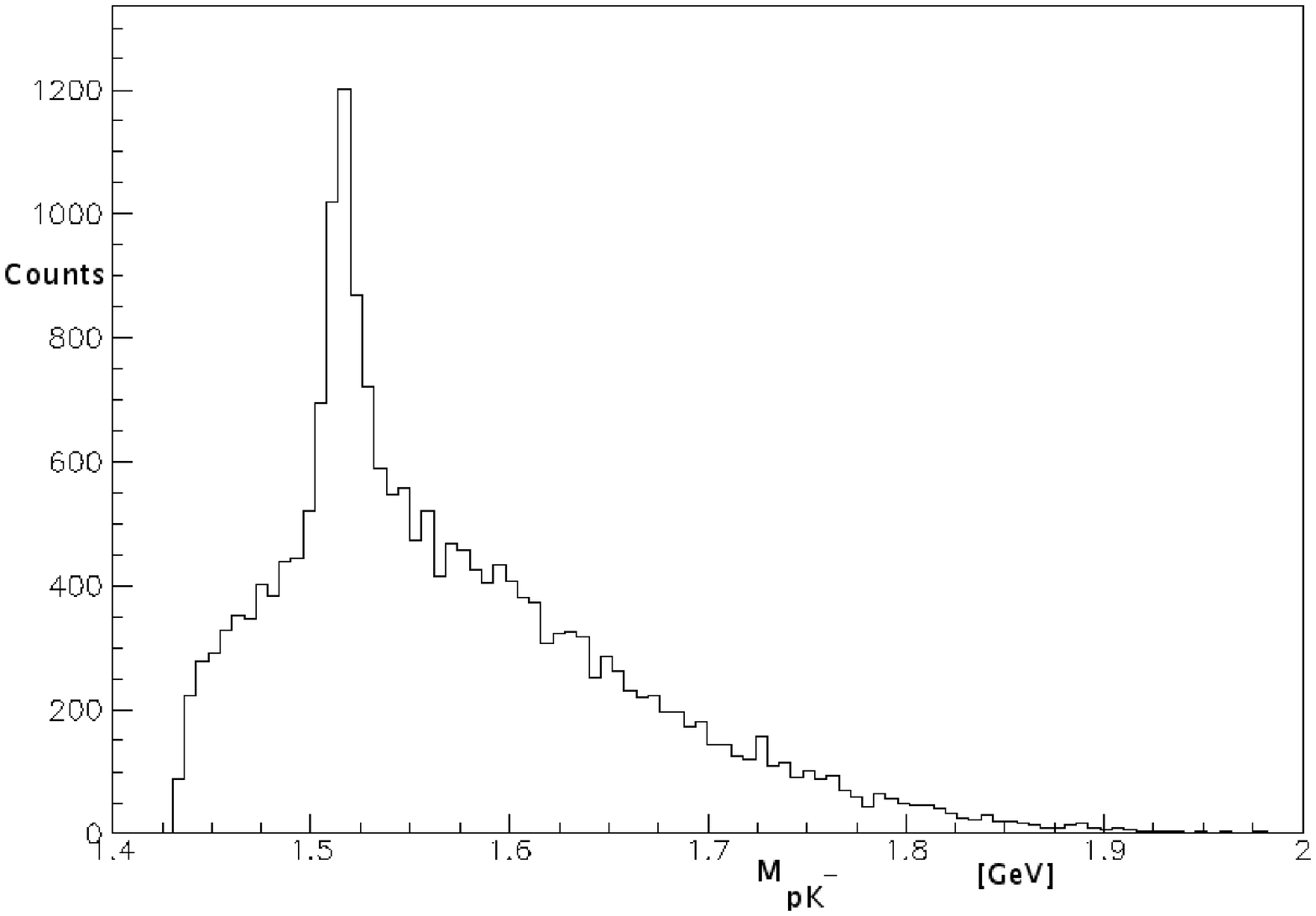}}
}&
\resizebox{0.5\textwidth}{!}{%
  \rotatebox{0}{\includegraphics{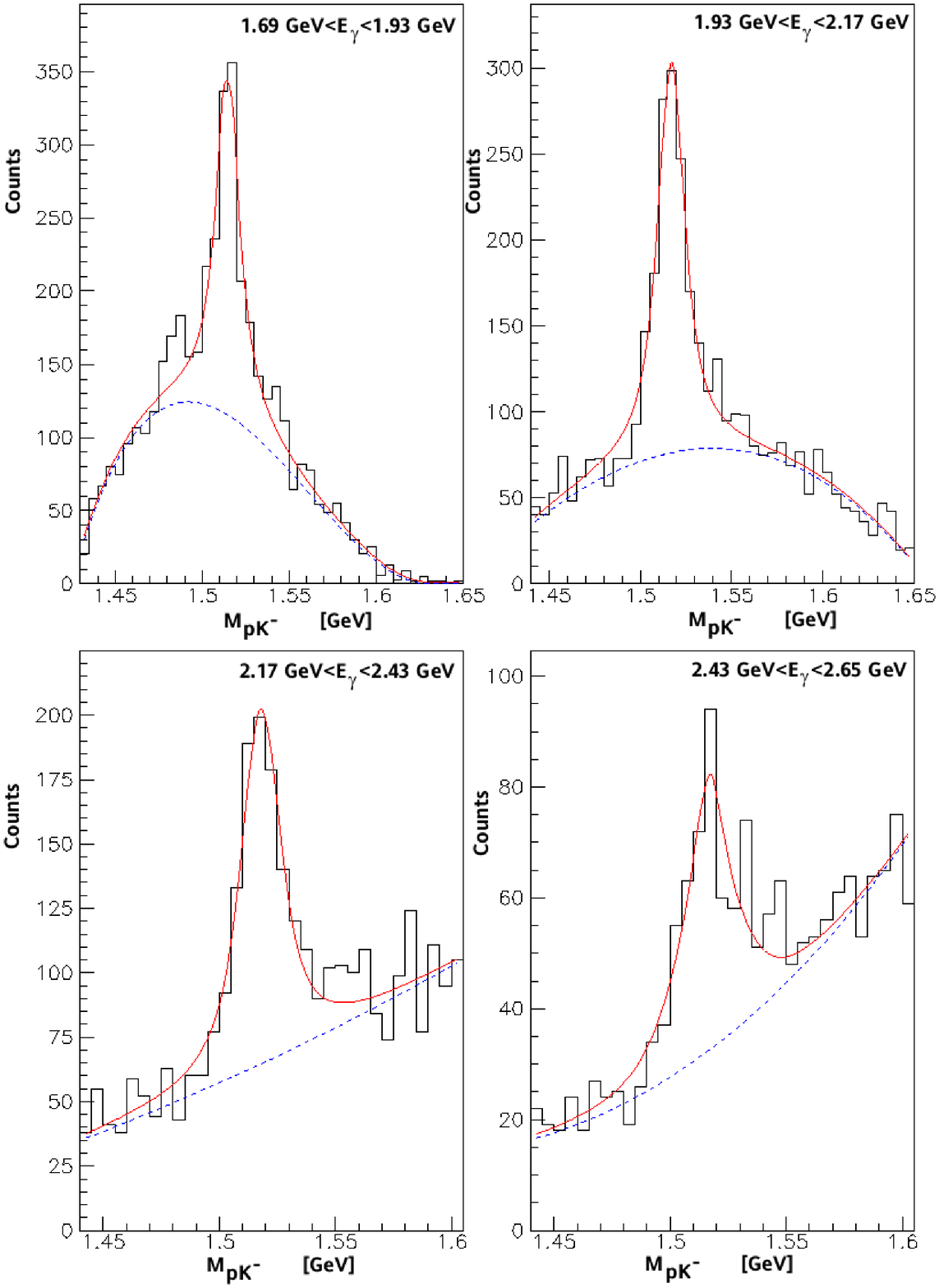}}
}\\
\end{tabular}
\caption{Reaction $\gamma p \rightarrow p K^+ K^-$: Invariant mass
  distribution of the $pK^-$ system with the peak of the $\Lambda(1520)$
  (left) and invariant mass distribution of the $pK^-$ system in four energy intervals (right panel).}
\label{pic:im_pk}
\end{figure*}

Figure \ref{pic:im_pk} shows the $pK^-$ invariant mass distributions
for the full photon energy range (left panel) and for four energy
bins (right panel). The distributions show a strong $\Lambda(1520)$
signal. The $\Lambda(1520)$ contribution was separated from the
background by fitting a polynomial of fifth order together with an
appropriate signal function. For this signal function a convolution
of a Breit-Wigner shape with natural decay width and of two
Gaussians, reflecting the experimental resolution (so-called double
Voigt function), was used. These fits were carried out with mass and
width of $\Lambda(1520)$ fixed to PDG values \cite{Amsler:2008zzb}.

\subsection{\boldmath $\Lambda(1520) \rightarrow n K^0_s$ \unboldmath}
\begin{figure*}[ht!]
\hspace{0.5cm}
\resizebox{0.9\textwidth}{!}{%
\rotatebox{0}{\includegraphics{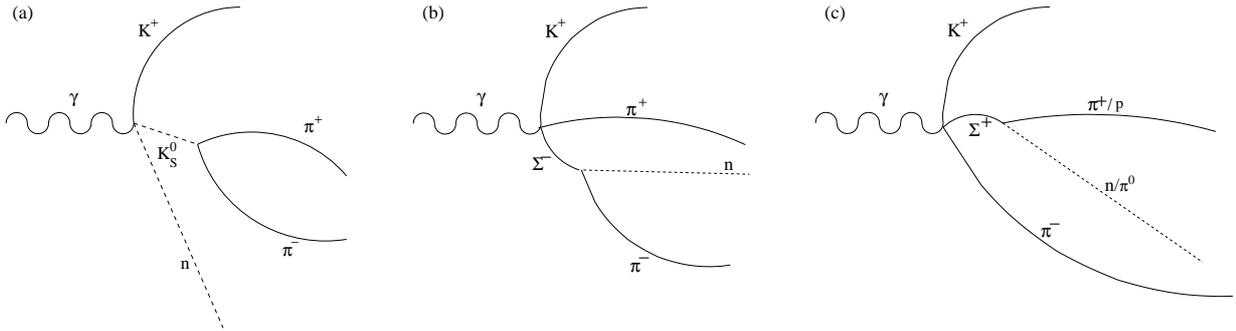}}
}
\caption{Topology of the reactions $\gamma p \rightarrow K^+ K^0_S n$ (a),
$\gamma p \rightarrow K^+ \Sigma^- \pi^+$ (b) and $\gamma p \rightarrow
  K^+ \Sigma^+ \pi^-$ (c).}
\label{Topology1}
\end{figure*}

\begin{table}[ht!]
\caption{\label{tab:reactions}Reaction hypotheses used for kinematic
fitting.}
\begin{center}
\renewcommand{\arraystretch}{1.1}
\begin{tabular}{cl}\\\hline\hline
 Hypothesis & Reaction \\ \hline
1 & $\gamma p \rightarrow p K^+ K^-$\\
2 & $\gamma p \rightarrow K^+ K^0_S n$\\
3 & $\gamma p \rightarrow K^+ \Sigma^+ \pi^-$\\
4 & $\gamma p \rightarrow K^+ \Sigma^- \pi^+$\\
5 & $\gamma p \rightarrow p \pi^+ \pi^-$\\
6 & $\gamma p \rightarrow p \pi^+ \pi^- \pi^0$\\
7 & $\gamma p \rightarrow n \pi^+ \pi^- \pi^+$\\
8 & $\gamma p \rightarrow p K^0_S K^0_L$\\
9 & $\gamma p \rightarrow K^0_S \Sigma^+$\\
10 & $\gamma p \rightarrow K^+ \Lambda \pi^0$\\
11 & $\gamma p \rightarrow K^0_S \Lambda \pi^+$\\
12 & $\gamma p \rightarrow K^0_S \Sigma^+ \pi^0$\\
13 & $\gamma p \rightarrow K^+ \Lambda$ \\
14 & $\gamma p \rightarrow K^+ \Sigma^0 $\\
15 & $\gamma p \rightarrow K^0_L \Lambda \pi^+$ \\
\hline\hline
\end{tabular}
\renewcommand{\arraystretch}{1.0}\end{center}
\end{table}
The topology of the reaction $\gamma p \rightarrow K^+ nK^0_s$,
where $K^0_s$ decays into $\pi^+\pi^-$, is shown in fig.
\ref{Topology1}(a). The secondary vertex of the $K^0_s$ decay was
reconstructed by a fit. Vertices with a total charge of zero and an
invariant mass near the $K^0_s$ mass ($|M_{\pi^+\pi^-} -
M_{K^0_s}|<0.03$\,GeV) were selected. Using time-of-flight
information it was checked that the mass of the positive
decay-particle is compatible with the pion mass. In addition it was
required that the mass of the second positively charged particle is
compatible with the mass of the $K^+$, in the mass range from
0.4\,GeV to 0.7\,GeV. The primary vertex was determined by a vertex
fit including the $K^+$ track and the line of flight of the $K^0_s$.
Events with a neutron in the final state were selected by a cut on
the missing mass recoiling against the $K^+K^0_s$ system
($|M_\mathrm{miss}-M_n|<0.07$\,GeV).

Finally  a kinematic fit was used to distinguish between the wanted
reaction and other channels (see table \ref{tab:reactions}).
Remaining background from the reactions $\gamma p \rightarrow K^+
\Sigma^\pm \pi^\mp$, which can populate the same final state, was
removed by
a cut on the invariant $n\pi^\pm$ mass in the range 1.180\,GeV $<
M_{n\pi^+} < $ 1.210\,GeV and 1.180 $< M_{n\pi^-} < $ 1.204\,GeV,
respectively.

\begin{figure*}[pt]
\hspace{0.5cm}
\resizebox{1.0\textwidth}{!}{%
  \begin{tabular} {c c}
    \rotatebox{0}{\includegraphics{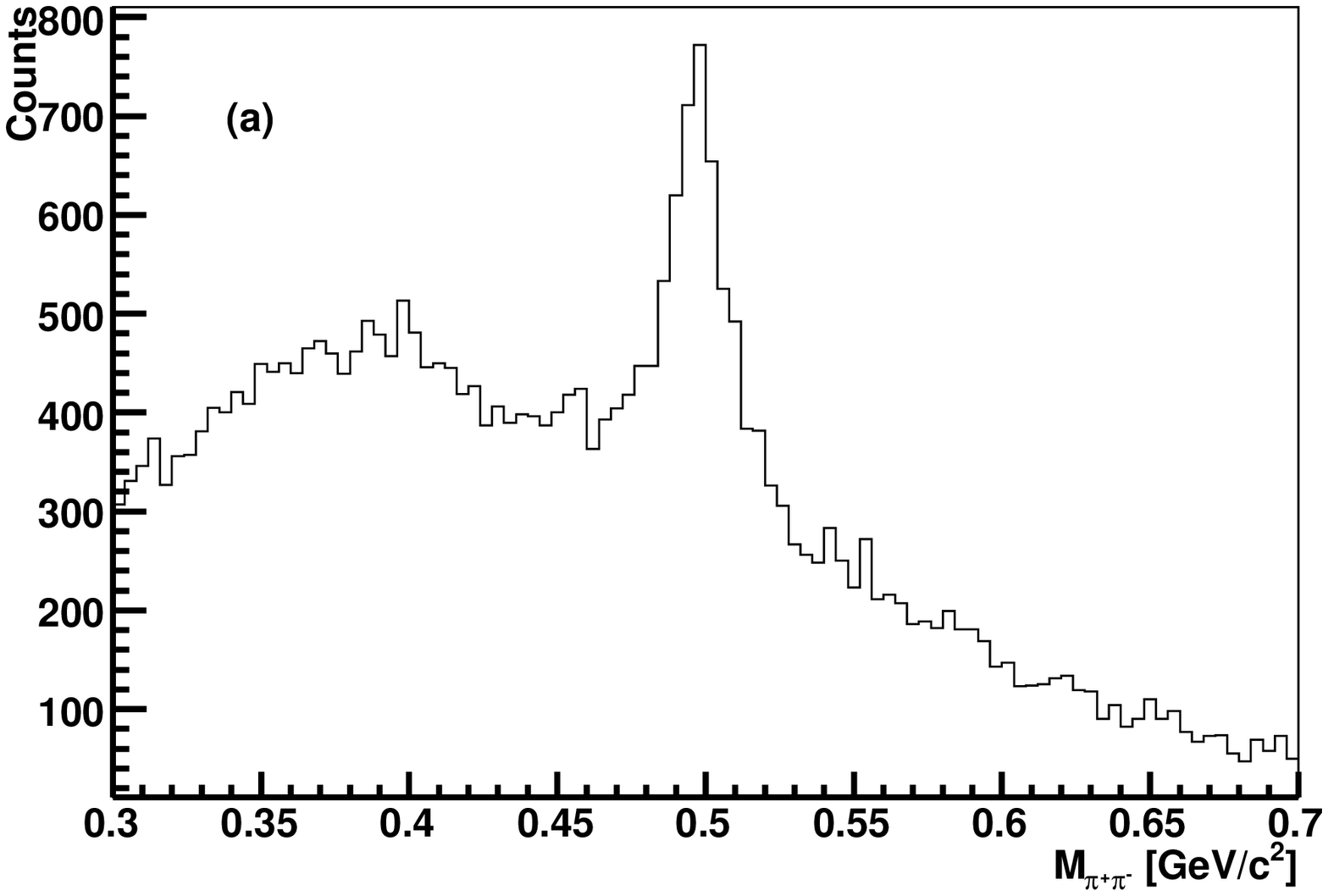}}\hspace{1.0cm}&
    \rotatebox{0}{\includegraphics{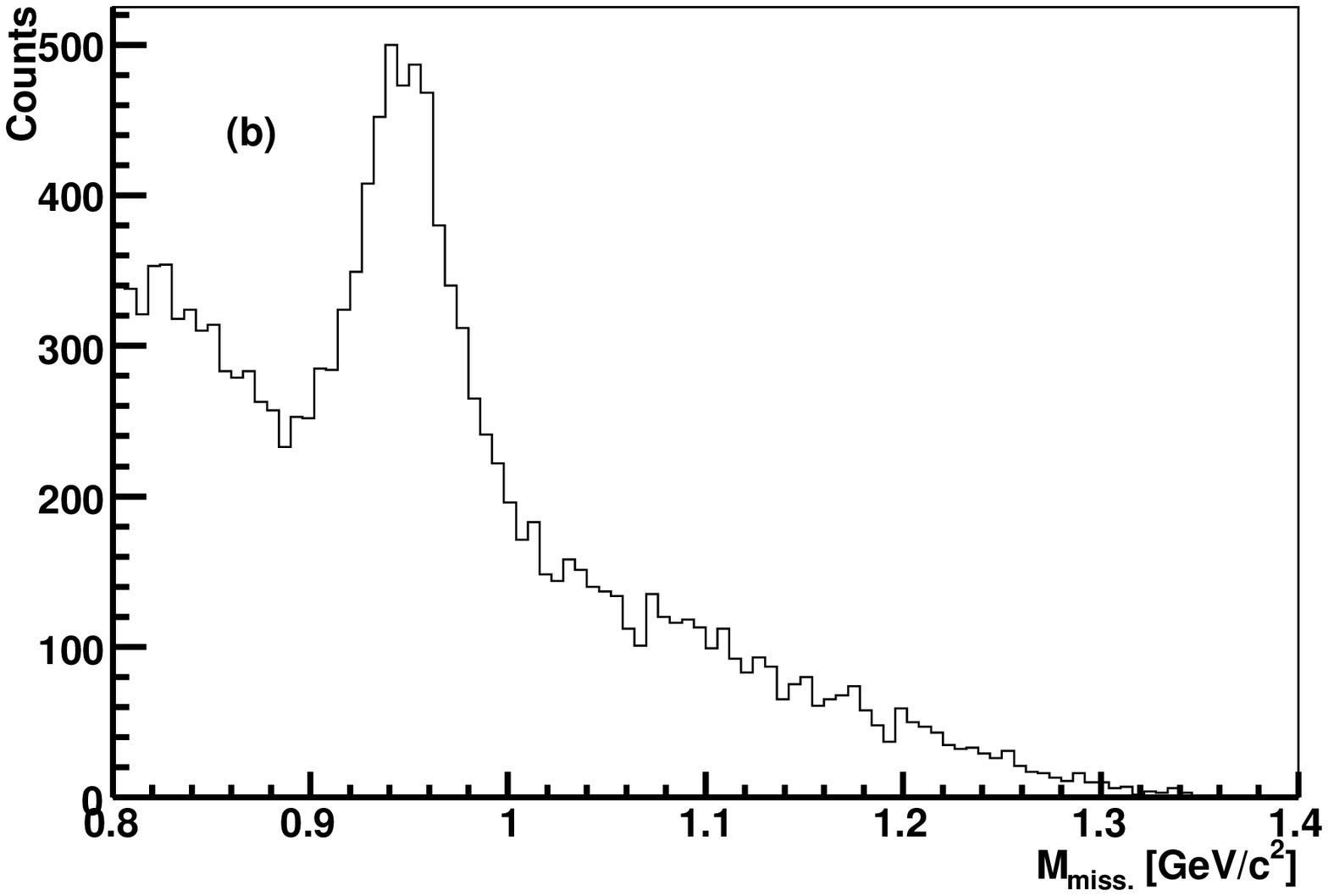}}\hspace{1.0cm}\vspace{-2mm}\\
  \end{tabular}
  }
\caption{Mass distributions for the reaction $\gamma p \rightarrow K^+K^0_sn$,
  (a) Invariant mass of $\pi^+$-$\pi^-$ pairs in preselected 3-track data with
  $K^0_s$-peak. The following cuts were applied: for the neutron
  $|M_\mathrm{miss}-0.94$\,GeV$|<0.075$\,GeV and for the $K^+$ a TOF mass
  information of $m>0.3$\,GeV. (b) Distribution of missing mass recoiling against
  the three tracks for all selected 3-track events showing a peak due to the
  missing neutron. \vspace{-2mm}}
\label{pic:mm_kplus_kplk0n}
\end{figure*}

\begin{figure*}[pt]
\begin{tabular} {c c}
  \resizebox{0.48\textwidth}{!}{\includegraphics{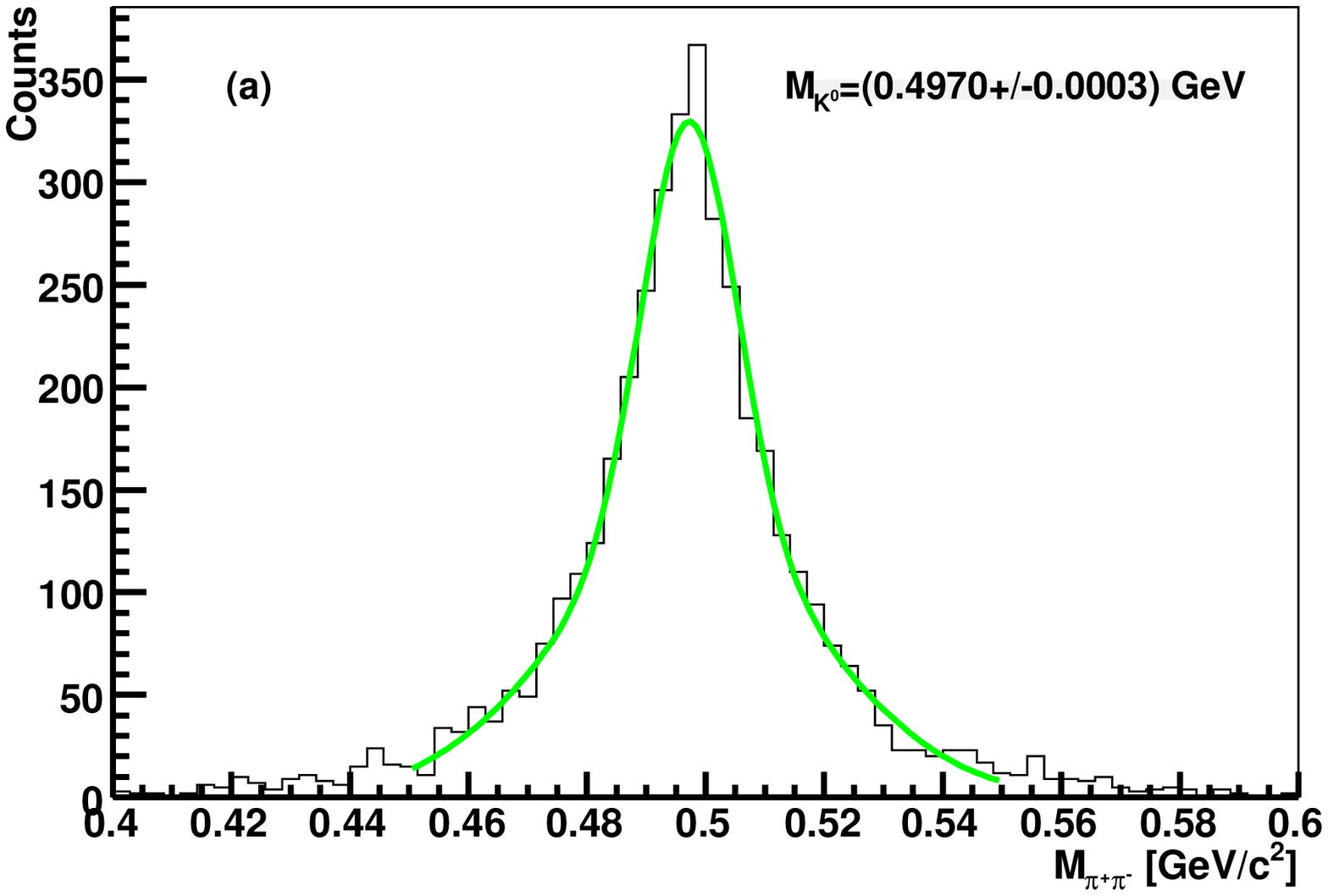}} &
  \resizebox{0.48\textwidth}{!}{\includegraphics{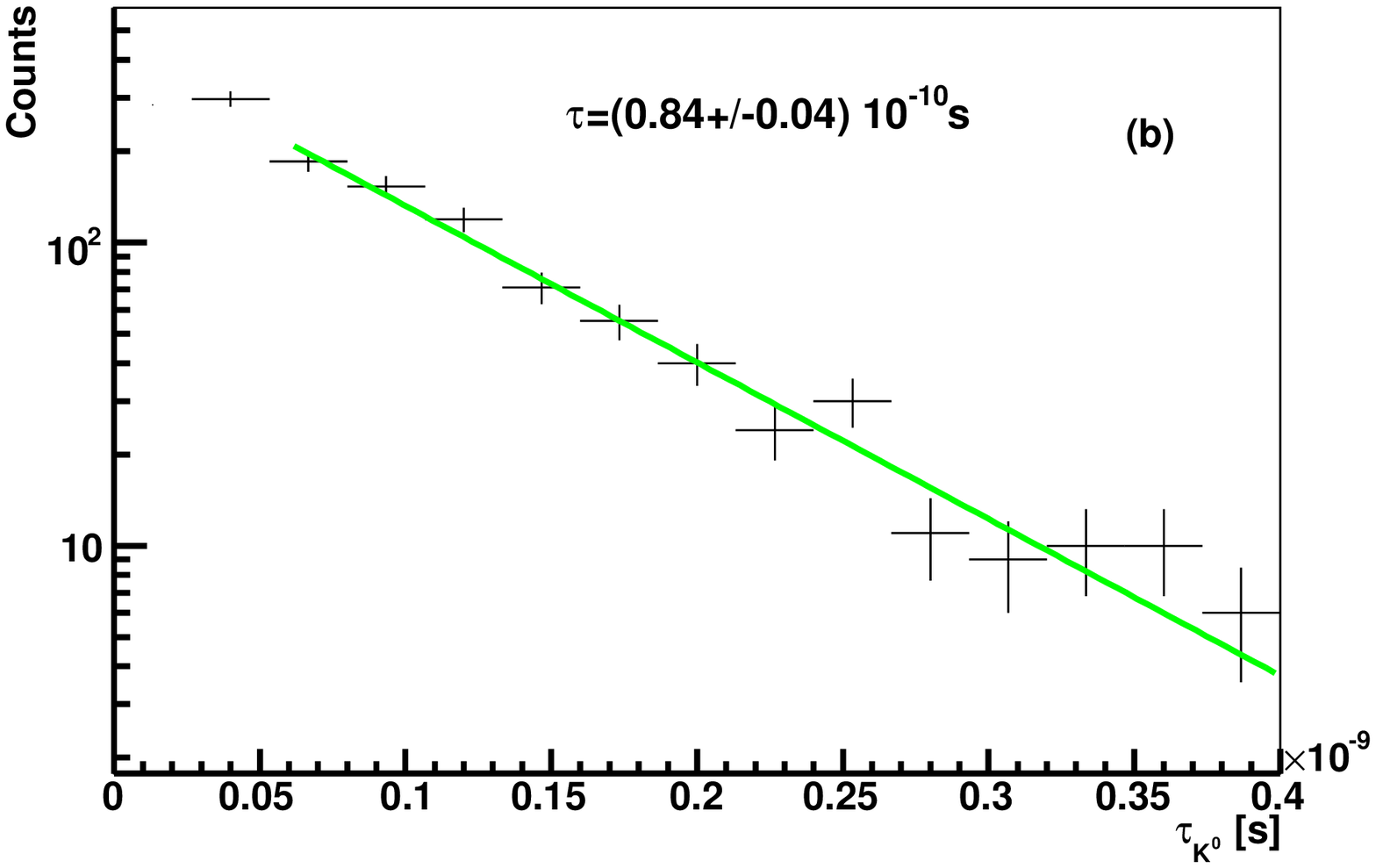}}\vspace{-2mm}\\
\end{tabular}
  \caption{(a) Invariant mass distribution of $\pi^+\pi^-$ pairs after complete selection
  procedure including a kinematical fit (see text), calculated from the
  measured 4-momenta and (b) decay time distribution of the
  $K^0_S$  for selected events from the reaction $\gamma p
  \rightarrow K^+ K^0_S n$.\vspace{-2mm}}
\label{pic:im_K0_}
\end{figure*}

\begin{figure}[ph]
{\centering
  \resizebox{0.48\textwidth}{!}{\includegraphics{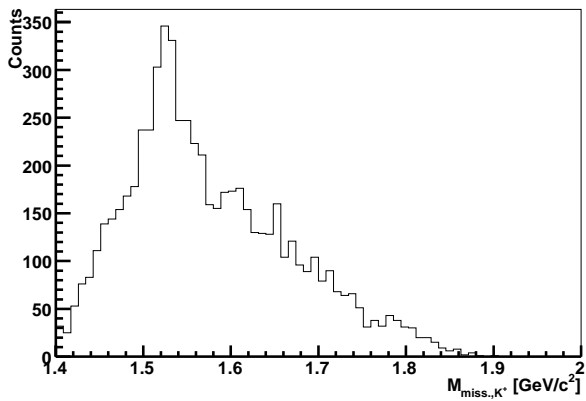}\vspace{-2mm}}
  \caption{\label{pic:im_K0_2}Missing mass recoiling against the $K^+$
  showing the $\Lambda(1520)$ peak.\vspace{-8mm}}
}

\end{figure}

Figure \ref{pic:mm_kplus_kplk0n}(a) shows the invariant mass
distribution of $\pi^+\pi^-$ pairs with the $K^0_s$ peak, after
preselection of the events, fig. \ref{pic:mm_kplus_kplk0n}(b) the
missing mass distribution calculated for the unmeasured neutral
particles with the neutron peak. In fig. \ref{pic:im_K0_}, after the
final event selection, the following distributions are presented:
(a) the invariant mass distribution of $\pi^+\pi^-$ pairs from
$K^0_s$ decay, (b) the $K^0_s$ decay time distribution and, in
fig.\,\ref{pic:im_K0_2}, the $K^0_sn$ mass distribution with the
$\Lambda(1520)$ peak.

The $\pi^+\pi^-$ mass distribution in
fig.\,\ref{pic:mm_kplus_kplk0n}(a) was fitted assuming a Gaussian
shape, giving a $K^0$ mass of (0.497$\pm$0.0003) GeV in agreement
with the nominal value \cite{Amsler:2008zzb}. Also the decay time
distribution in fig.\,\ref{pic:im_K0_}(b) with
$\tau$=(0.84$\pm$0.04)$\cdot 10^{-10}$s is consistent with the
nominal value. Similar to the $pK^-$ case, the $\Lambda(1520)$ peak
was fitted in the $nK^0_s$ mass distributions
in four photon-energy bins with a double Voigt function and a
background term (not shown).\\[-6ex]

\subsection{\boldmath $\Lambda(1520) \rightarrow \Sigma^\pm \pi^\mp$\unboldmath}
\begin{figure*}[ht!]
\centerline{
\resizebox{0.9\textwidth}{!}{%
\begin{tabular} {c c c}
  \rotatebox{0}{\includegraphics{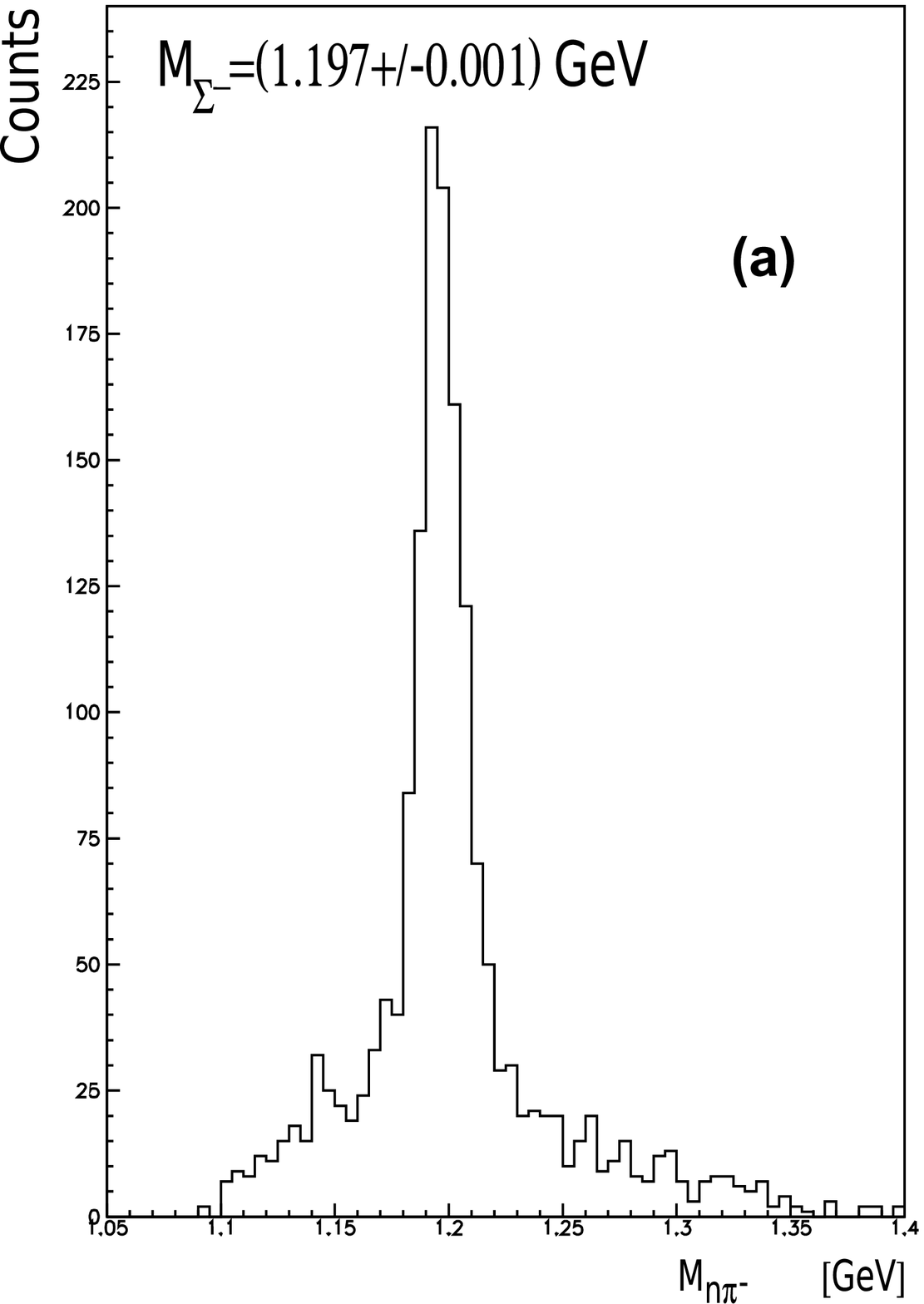}}\hspace{1.0cm}&
  \rotatebox{0}{\includegraphics{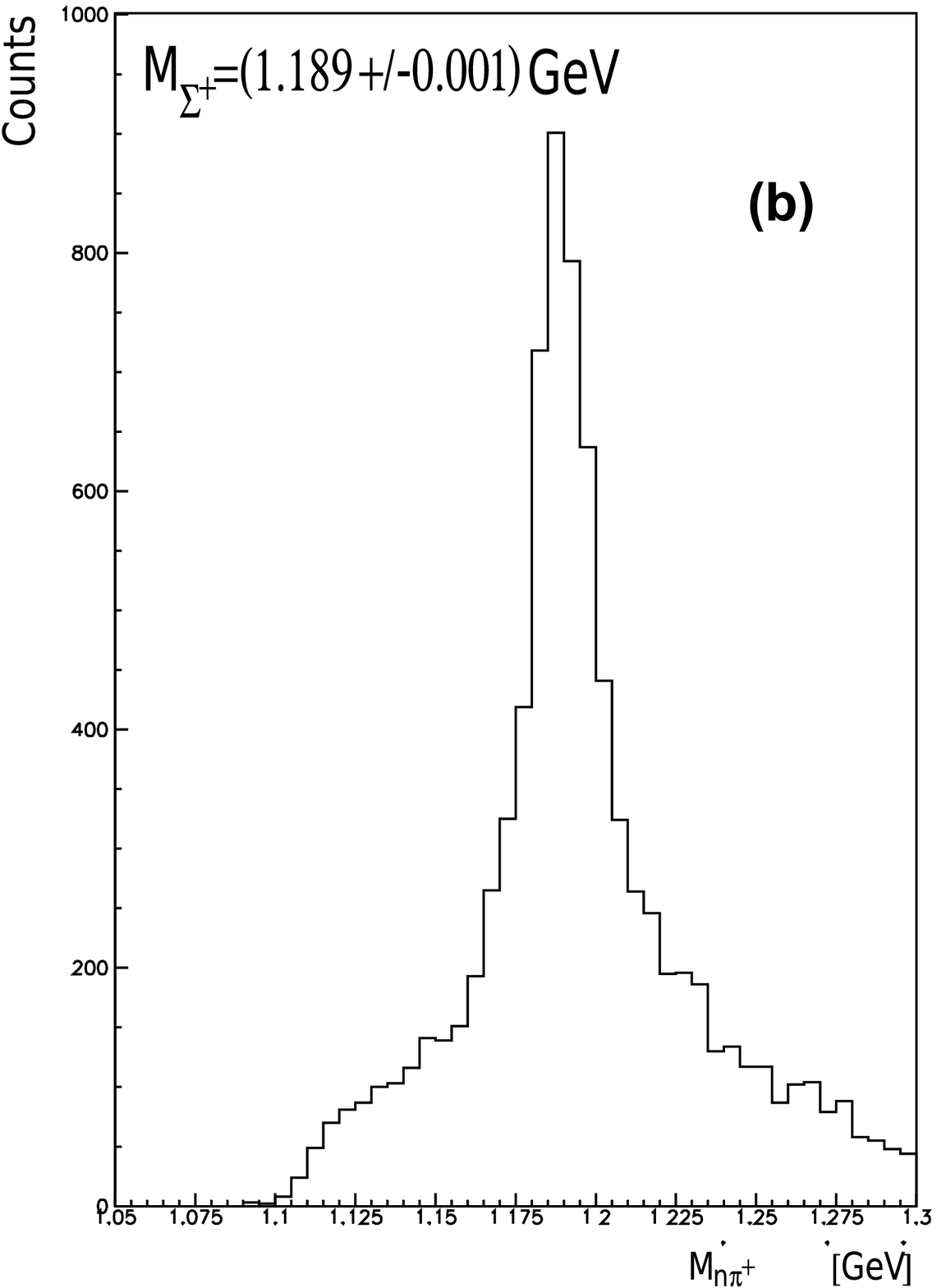}}\hspace{1.0cm}&
  \rotatebox{0}{\includegraphics{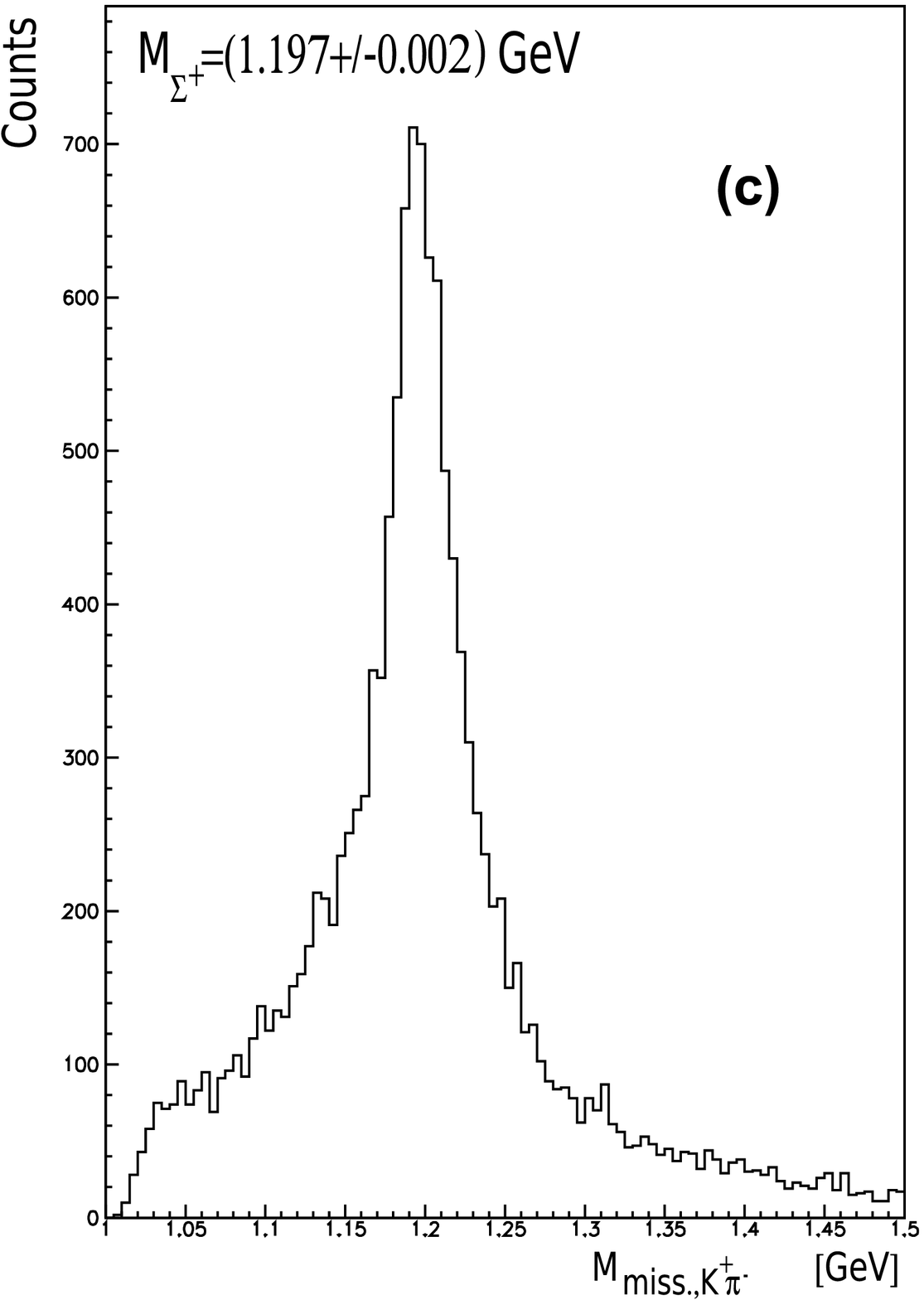}}\\
\end{tabular}
}} \caption{(a) Invariant mass distribution $M_{n\pi^-}$ of the
$n\pi^-$ system of the $\Sigma^-$-decay for selected events from the
reaction $\gamma p \rightarrow K^+ \Sigma^- \pi^+$, (b) Invariant
mass $M_{n\pi^+}$ of the $n\pi^+$ system of the $\Sigma^+$-decay for
selected events from the reaction $\gamma p \rightarrow K^+ \Sigma^+
\pi^-$, (c) Missing mass $M_{\mathrm{miss.},K^+\pi^-}$ recoiling
against the $K^+\pi^-$ system for the decay-channel $\Sigma^+
\rightarrow p \pi^0$.}
\label{pic:imsig1}
\end{figure*}
%die Bilder waren vorher zusammen
\begin{figure*}[ht!]
\centerline{
\resizebox{0.9\textwidth}{!}{%
\begin{tabular}{c c}
  \rotatebox{0}{\includegraphics{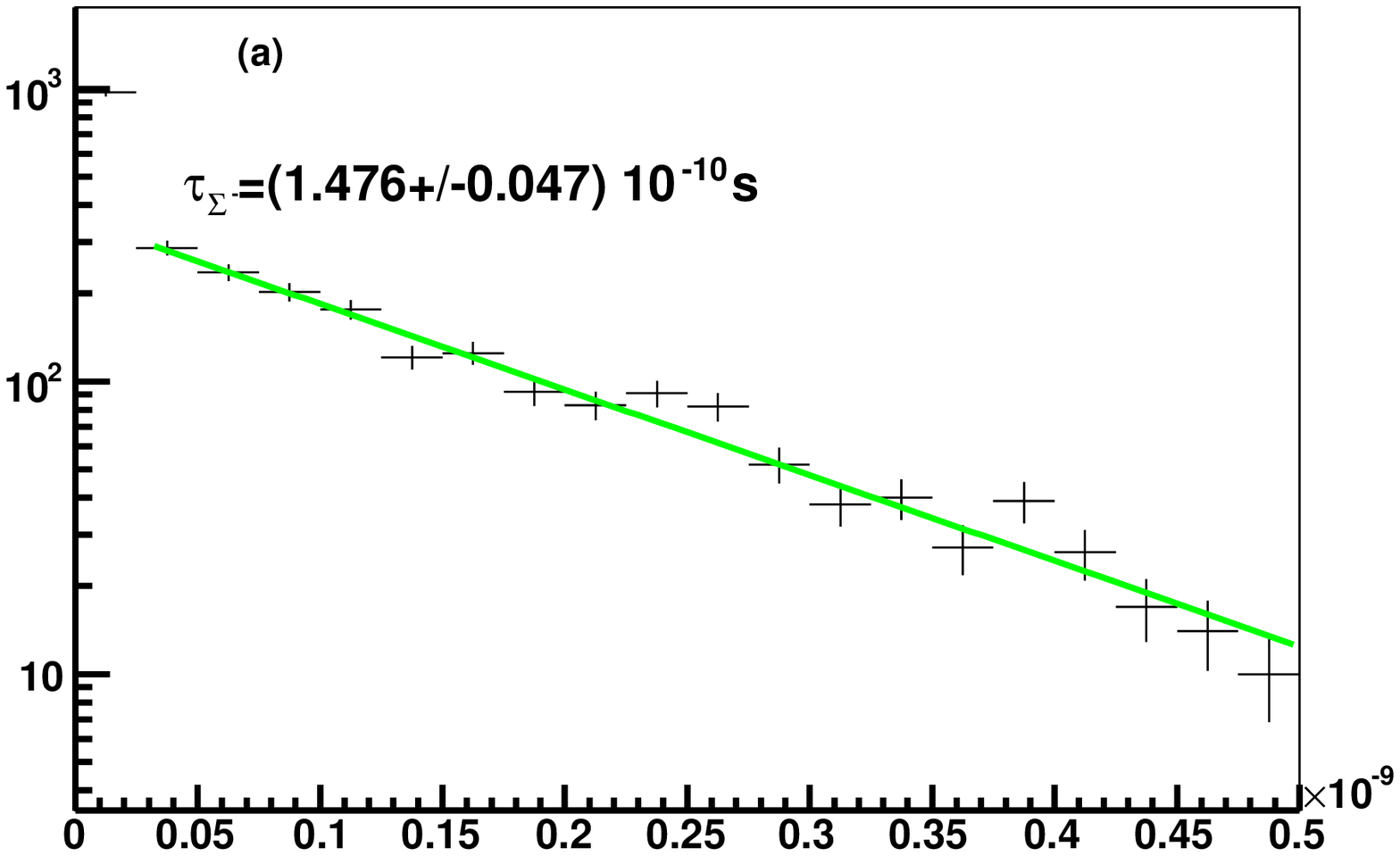}}\hspace{1.0cm}&
  \rotatebox{0}{\includegraphics{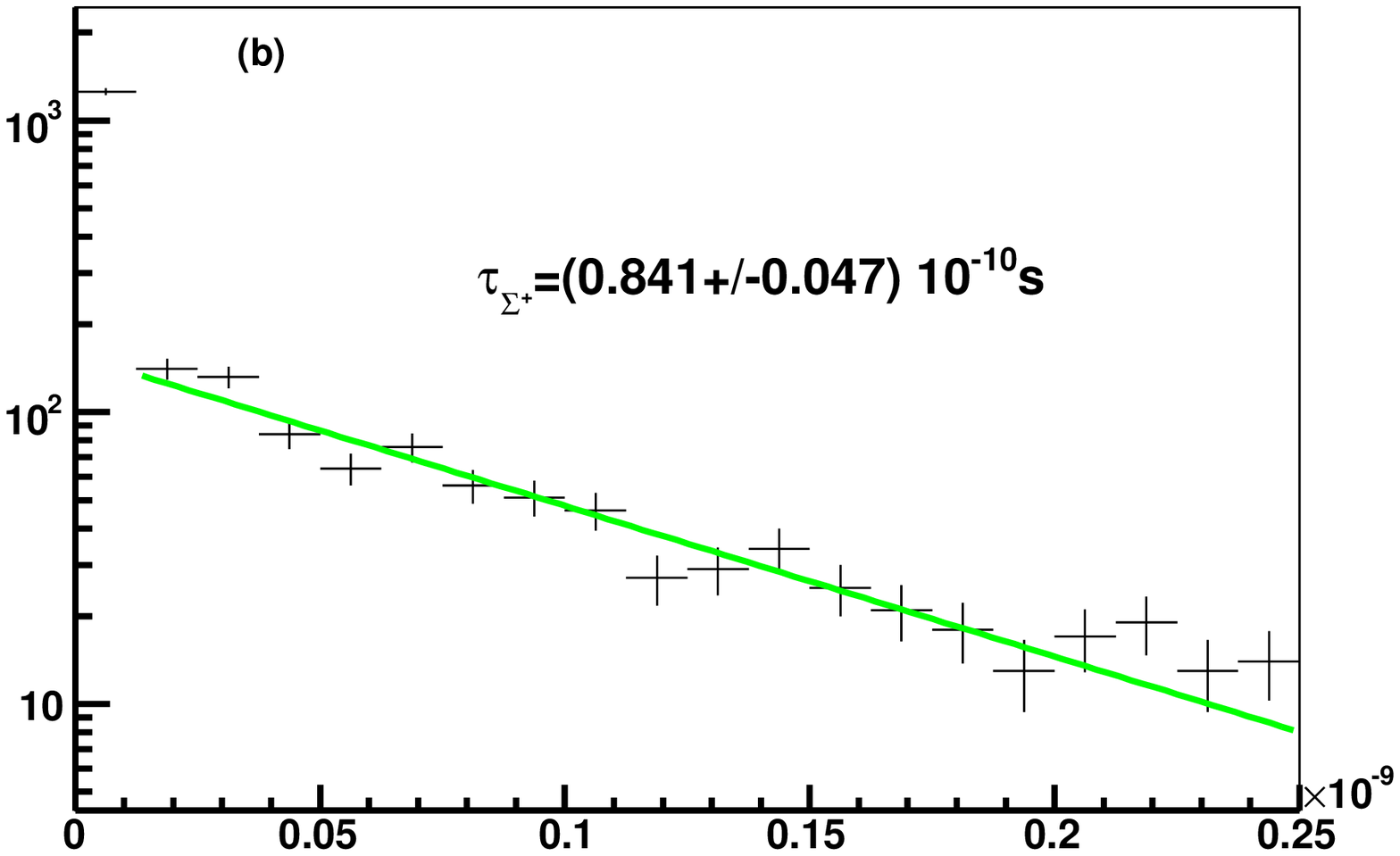}}\hspace{1.0cm}\\
\end{tabular}
}} \caption{(a) Decay time distribution of $\Sigma^-$, (b) Decay
time distribution of $\Sigma^+$.}
\label{pic:imsig2}
\end{figure*}

The topologies of the reactions $\gamma p \rightarrow
K^+\Sigma^-\pi^+$ with $\Sigma^-$ decaying into $n\pi^-$ and $\gamma
p \rightarrow K^+\Sigma^+\pi^-$ with $\Sigma^+$ into $n\pi^+$ and
$p\pi^0$ are shown in figs. \ref{Topology1} (b) and (c),
respectively. First, the primary vertex was reconstructed in a fit
of the intersection of tracks trying all possible pairs of charged
particles. The vertex fit with the best fit probability was
accepted. Then the $\Sigma^\pm$ momentum was reconstructed at this
vertex by a kinematic fit to hypotheses 3 and 4 (table
\ref{tab:reactions}). As next step, the  decay vertex of
$\Sigma^\pm$  was  reconstructed in a fit as intersection of the
reconstructed $\Sigma$ track  with the third charged track. The
$\Sigma^\pm$ decay hypotheses were tested by a kinematic fit at the
decay vertex. The $K^+$ mass, determined from time-of-flight
measurement had to lie in the region $0.4 < M_{K^+} < 0.7$\,GeV. In
order to remove background from other reactions, further kinematic
cuts were applied. The masses of $n$ and $\pi^0$ candidates,
calculated as missing mass at the $\Sigma$ decay vertex, had to be
in the regions $|M_\mathrm{miss} - M_n|<0.07$\,GeV and
$|M^2_\mathrm{miss} - M^2_{\pi^0}| < 0.005$\,GeV$^2$, respectively.
The invariant mass of the $n\pi^\mp$ system, calculated at the decay
vertex, had to be in the range of the nominal mass of the $\Sigma$
($|M_{n\pi^\pm}-M_{\Sigma^{\pm}}|<0.017$\,GeV). For $\Sigma^+
\rightarrow p\pi^0$, it was requested that the missing mass
recoiling against the $K^+\pi^-$ system at the primary vertex is in
the range $1.165 < M_{\mathrm{miss}K^+\pi^-} < 1.229$\,GeV.
Remaining background from the reaction $\gamma p \rightarrow
nK^+K^0_s$ was removed by requiring $|M_{\pi^+\pi^-} - M_{K^0_s}| >
0.03$\,GeV. In addition it was demanded that the $K^+$ mass, as
determined from time-of-flight measurement, was in the $0.4 <
M_{K^+} < 0.7$\,GeV range.

\begin{figure*}[ht!]
\hspace{0.5cm}
\resizebox{0.9\textwidth}{!}{%
\begin{tabular} {cc}
\includegraphics{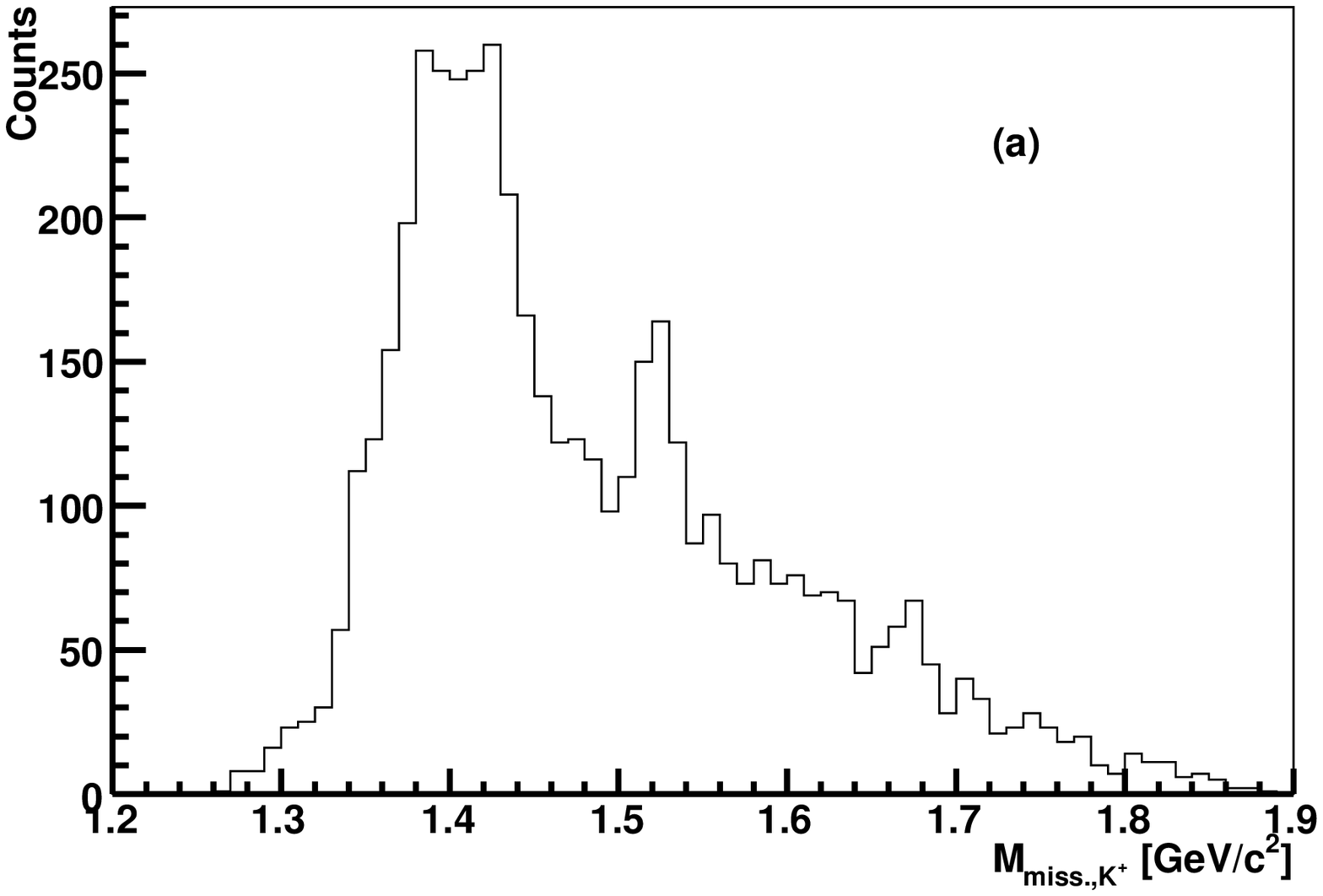}\hspace{1.0cm}\vspace{-142mm}&\\
&\includegraphics{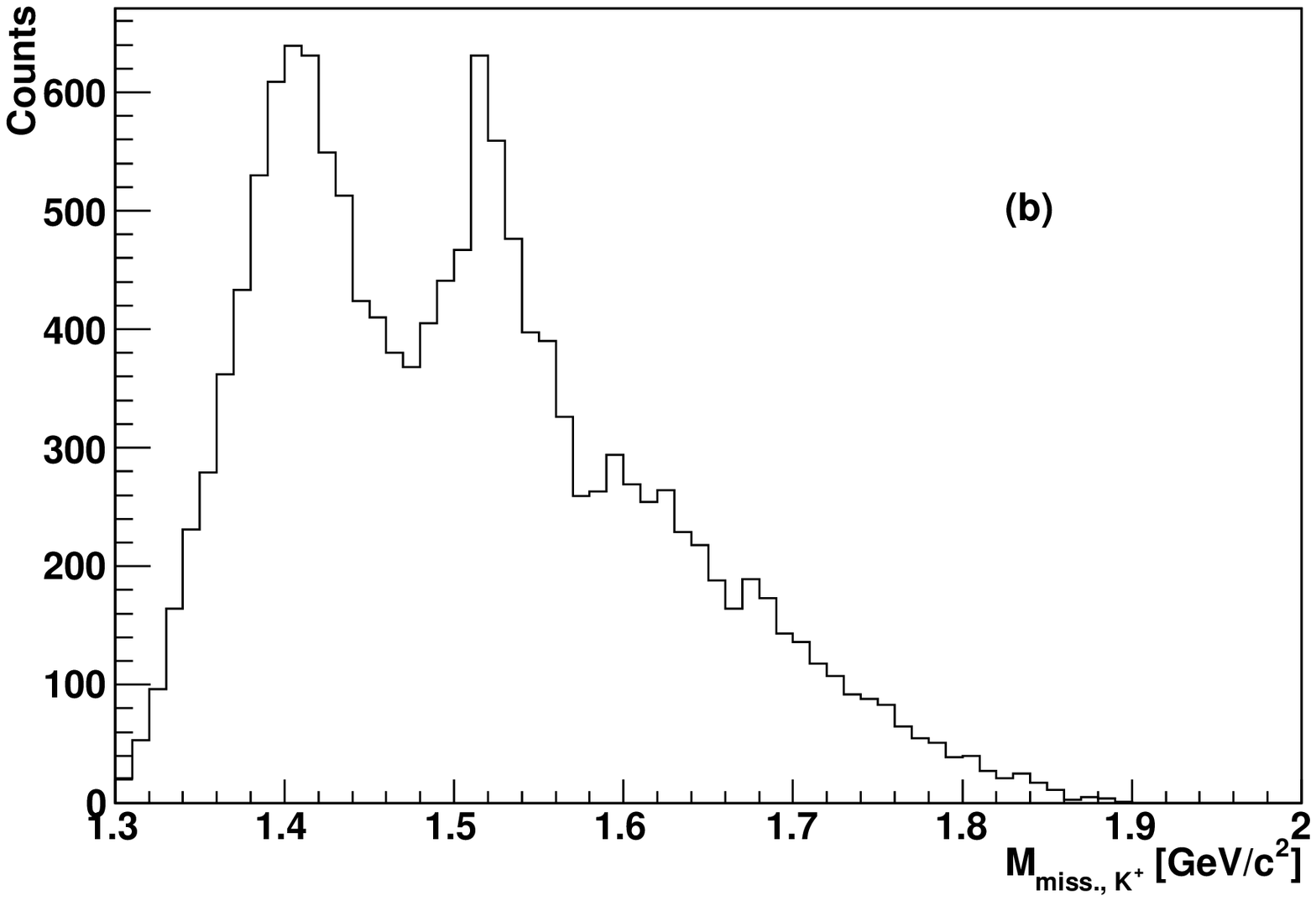}\hspace{1.0cm}\\
\end{tabular}
} \caption{(a) Missing mass distribution $M_{miss.,K^+}$ recoiling
   against the $K^+$ for the reaction $\gamma p \rightarrow K^+
   \Sigma^- \pi^+$ with the peak of
  the $\Lambda(1520)$ and a peak due to the $\Sigma(1385)/\Lambda(1405)$ complex,
  (b) Missing mass distribution $M_{miss.,K^+}$ for the reaction $\gamma p
  \rightarrow K^+ \Sigma^+ \pi^-$ (for both decay-channels of the $\Sigma^+$)
  with peaks of the $\Lambda(1520)$ and $\Sigma(1385)/\Lambda(1405)$.}
  \label{pic:IM_Lambda1520_Sig_channels}
\end{figure*}

For the decay $\Sigma^+ \rightarrow p\pi^0$, background events from
the reaction $\gamma p \rightarrow K^+\Lambda\pi^0$  with
$\Lambda\rightarrow p\pi^-$ were removed by an anti-cut around the
$\Lambda$ mass ($1.098 < M_{p\pi^-} < 1.128$\,GeV). Figure
\ref{pic:imsig1} shows the invariant mass distributions of (a)
$n\pi^-$ from the $\Sigma^-$ decay, (b) $n\pi^+$ from the $\Sigma^+$
decay, and (c) the missing mass recoiling against the $K^+\pi^-$
system for the decay channel $\Sigma^+\rightarrow p\pi^0$. Figure
\ref{pic:imsig2} presents decay time distributions of $\Sigma^-$ (a)
and $\Sigma^+$ (b) after the complete event selection and
kinematical fits. The mass and lifetime values are consistent with
the PDG data.

Figure \ref{pic:IM_Lambda1520_Sig_channels} shows the distributions
of missing masses recoiling against the $K^+$ for the reactions
$\gamma p \rightarrow K^+\Sigma^-\pi^+$ (a) and  $\gamma p
\rightarrow K^+\Sigma^+\pi^-$ (b). Again the distributions were
fitted in four photon energy bins with a double Voigt function and a
polynomial background term.

\subsection{\boldmath $\Lambda(1520) \rightarrow \Lambda\pi^+\pi^-$
  \unboldmath}

\begin{figure}[pt]
{\centering
\resizebox{0.44\textwidth}{!}{\includegraphics{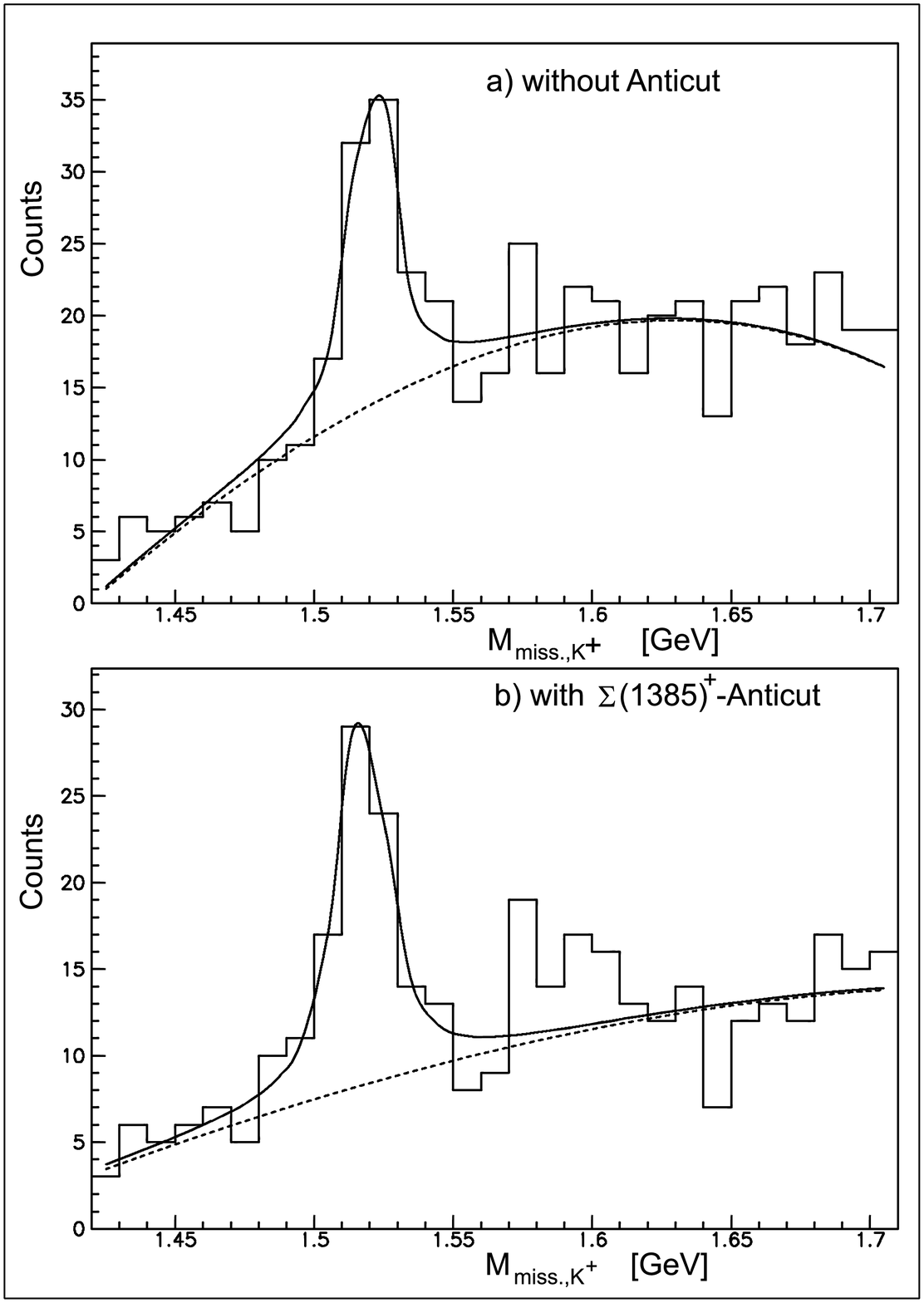}}}
\caption{Decay of $\Lambda(1520)$-hyperons in the channel
$\Lambda(1520) \rightarrow \Lambda \pi^+ \pi^-$. Missing-mass
distribution $M_{miss.,K^+}$ with the peak of the
$\Lambda(1520)$-hyperon a) without further cuts, b) with an anticut
in the region of the missing mass $M_{miss.,K^+\pi^-}$ of
$\Sigma(1385)^+$ decays. Full lines describe the fit, dashed lines
the background assumption. \vspace{-4mm}}
\label{pic:lam1520Sigmaminuscut1}
\end{figure}
\begin{figure}[pt]
\resizebox{0.44\textwidth}{!}{\includegraphics{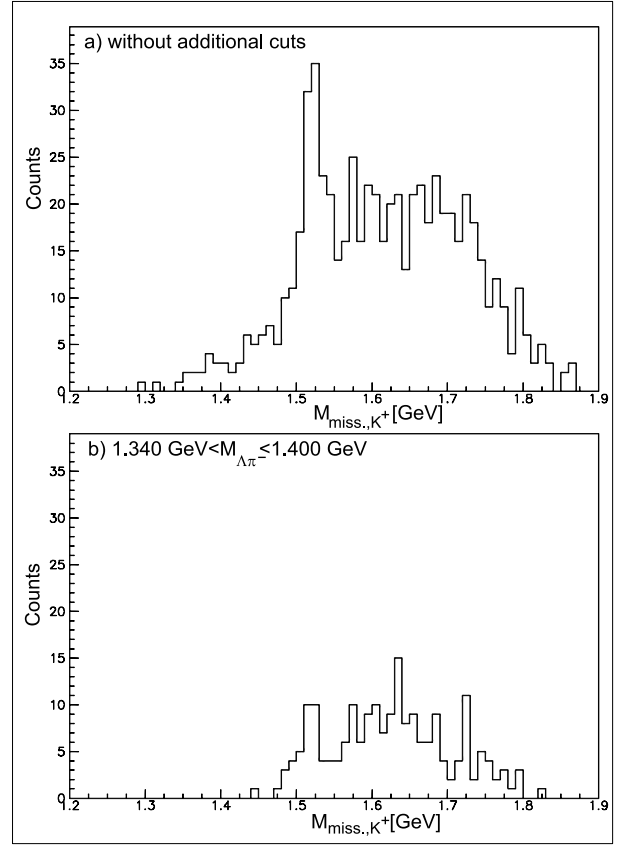}}
%\end{tabular}
\caption{Decay of $\Lambda(1520)$-hyperons in the channel
$\Lambda(1520) \rightarrow \Lambda \pi^+ \pi^-$. Missing-mass
distribution $M_{miss.,K^+}$ with the peak of the
$\Lambda(1520)$-hyperon a) without further cuts, b) Missing mass
plot of the $\Lambda(1520)$-hyperons with an invariant mass
$M_{\Lambda \pi^+}$ compatible with the decay channel $\Lambda(1520)
\rightarrow \Sigma(1385)^+\pi^-$.\vspace{-4mm}}
\label{pic:lam1520Sigmaminuscut2}
\end{figure}

Events from the reaction $\gamma p \rightarrow
K^+\Lambda\pi^+\pi^-$, where $\Lambda$ decays into $p\pi^-$, were
reconstructed from the photon energy and five charged tracks. The
decay vertex of the $\Lambda$ candidate was tentatively
reconstructed as intersection of a positively charged track with
each of the negative tracks, accepting the fit with the calculated
invariant mass closest to the nominal $\Lambda$ mass. The primary
vertex was determined using the reconstructed $\Lambda$ line of
flight and the remaining tracks. To reduce background from other
reactions, several cuts were applied. The missing mass recoiling
against $K^+\pi^+\pi^-$ had to lie in the $\Lambda$ mass region
($1.075 < M_\mathrm{miss} < 1.149$\,GeV). The longitudinal momentum
had to be conserved to better than 0.15\,GeV. The $K^+$ mass
assignment obtained from the time-of-flight measurement had to be in
the range $0.4 < M_{K^+} < 0.7$\,GeV. Finally a kinematic fit was
used to select events of the wanted reaction. The distribution of
the invariant mass opposite to $K^+$ is shown in figs.
\ref{pic:lam1520Sigmaminuscut1} and \ref{pic:lam1520Sigmaminuscut2}.
The number of $\Lambda(1520) \rightarrow \Lambda\pi^+\pi^-$ was
determined as before.

\begin{figure*}[pt]
\hspace{0.5cm}
\resizebox{1.0\textwidth}{!}{%
\begin{tabular} {c c}
  \rotatebox{0}{\includegraphics{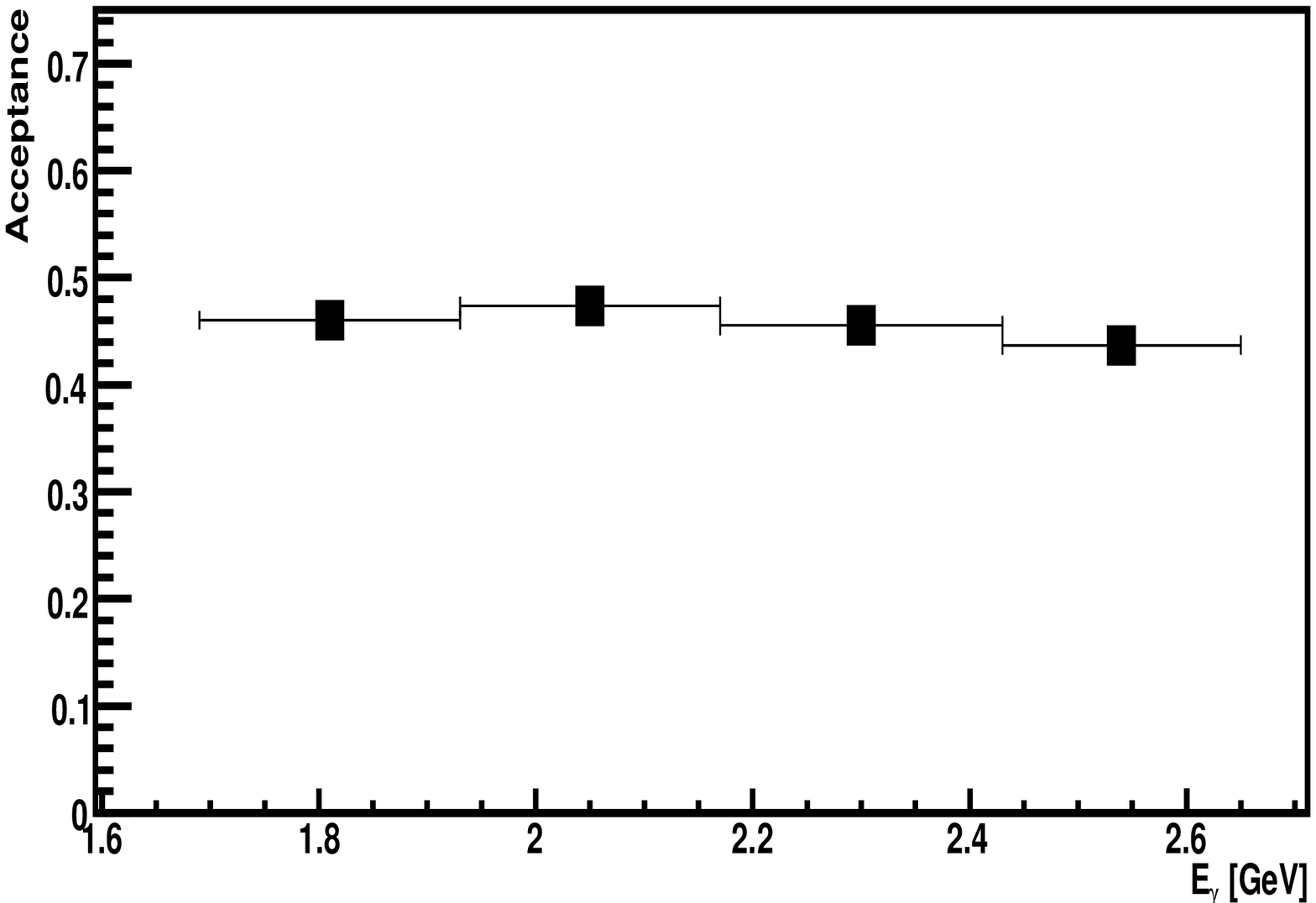}}\hspace{1.0cm}&
  \rotatebox{0}{\includegraphics{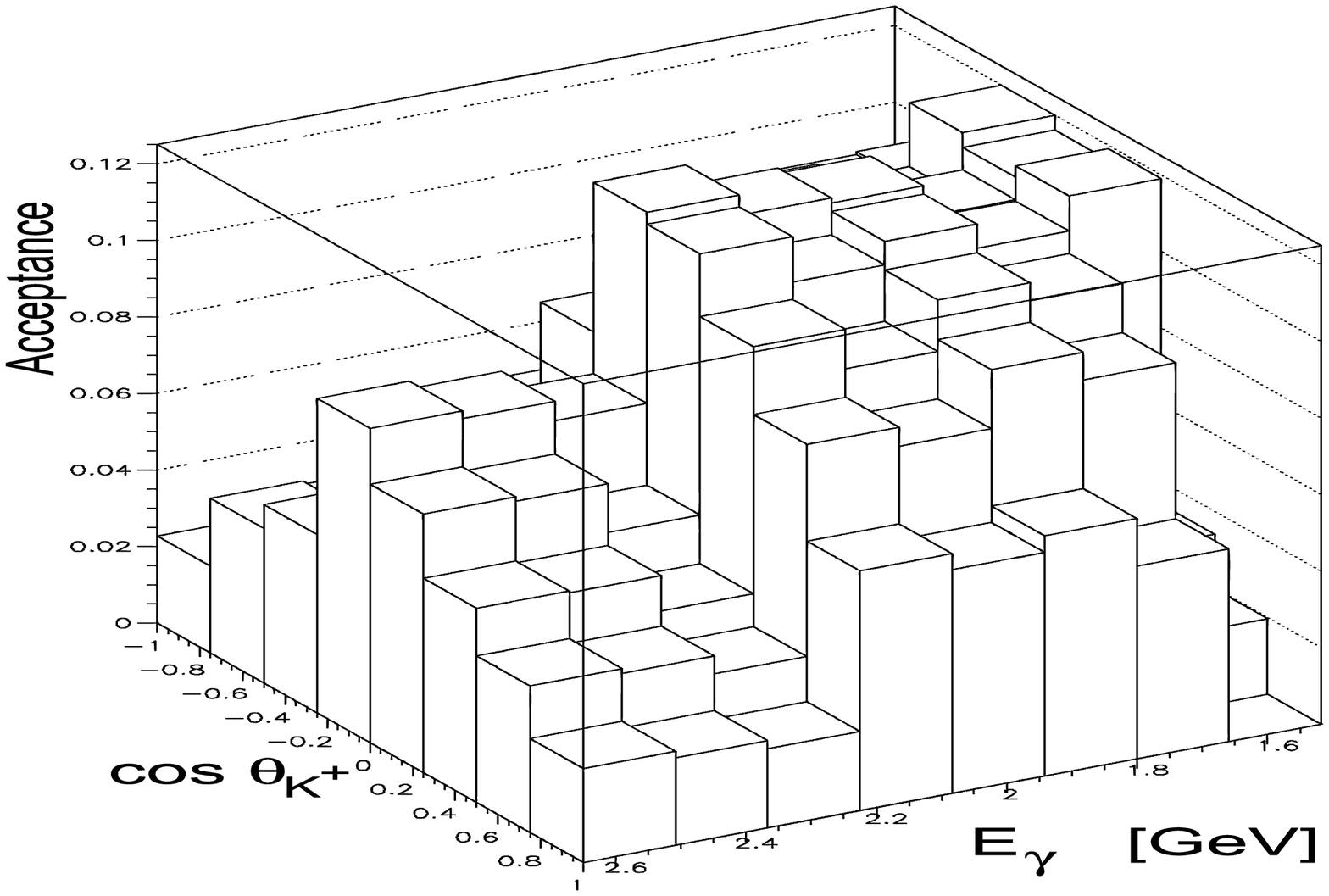}}\\
\end{tabular}
} \caption{Left panel: Acceptance of the reaction $\gamma p
\rightarrow K^+
  \Lambda(1520)$ summed over the decay   modes as a function of  $E_\gamma$.
  Right panel: Acceptance for the reaction $\gamma p \rightarrow K^+ K^0_s
  n$ vs. $E_{\gamma}$ and $\cos{\theta_{K^+}}$.
}
\label{pic:acceptance}
\end{figure*}

\section{Background from other reactions}
\label{sec:background}

In order to estimate remaining background contributions which pass
the selection process, background reactions (see table
\ref{tab:reactions}) were simulated according to phase space in the
photon energy range of the data, and processed in the same way as
real detector data. In addition to the reactions in table
\ref{tab:reactions}, $\gamma p \rightarrow p\pi^+\pi^-\pi^+\pi^-$
and $\gamma p \rightarrow p\pi^+\pi^-\pi^+\pi^-\pi^0$ were simulated
for the $\Lambda\pi^+\pi^-$ decay mode. The uncertainty in the shape
of the background yield a total systematic uncertainty in the
$\Lambda(1520)$ cross section of less than 10\% for all decay
channels.

\section{Acceptance}
\label{sec:acceptance}
The acceptance was determined as a function of the photon energy
$E_\gamma$ and $\cos\theta_{K^+}$ in the c.m.s. by means of Monte
Carlo simulations using the GEANT programme package \cite{GEANT}.
Events from the reaction $\gamma p \rightarrow K^+ \Lambda(1520)$
with $\Lambda(1520)$ decaying into the final states $pK^-$,
$nK^0_S$, and $\Sigma^{\pm}\pi^{\mp}$ were generated by the
SAGE-generator \cite{SAGE} with propagation of $K^0_S$ and
$\Sigma^{\pm}$ according to their lifetimes and the decays $K^0_S
\rightarrow \pi^+ \pi^-$, $\Sigma^{-} \rightarrow n\pi^{-}$,
$\Sigma^{+} \rightarrow p\pi^{0}$, and $\Sigma^{+} \rightarrow
n\pi^{+}$, respectively. Charged particles in the final states were
tracked through the drift chamber system of the SAPHIR detector
taking into account the magnetic field and multiple scattering in
all materials passed. Simulated events were analyzed in the same way
as real data. The determination of the acceptance comprised the
trigger efficiency of the data taking periods, the event
reconstruction efficiency and the efficiency of the selection
process. On average the acceptance was of the order of $44\%$ for
$\gamma p \rightarrow K^+ \Lambda(1520) \rightarrow K^+ p K^-$ and
$8\,\%$ for $\gamma p \rightarrow K^+ K^0_sn, K^+ \Sigma^{\pm}
\pi^{\mp}$. The reaction acceptance and, as an example, details for
the $K^+K^0_sn$ final state are given in fig. \ref{pic:acceptance}.

\section{Results}
\label{sec:crosssec}
\subsection{Total cross section for \boldmath $\gamma p
  \rightarrow K^+\Lambda(1520)$ \unboldmath}

The number of $\Lambda(1520)$ events was determined by fits to the
$pK^-$, $nK^0_s$, $\Sigma^-\pi^+$, and $\Sigma^+\pi^-$ mass
distributions in five $\cos\theta_{K^+}$ bins and for four bins in
the photon energy covering the range from 1.69\,GeV to\,2.65\,GeV.
The fits were carried out assuming a Breit-Wigner form with a decay
width fixed to the nominal value. Whilst in the $nK^0_s$ and $pK^-$
mass distributions only the $\Lambda(1520)$ provided a narrow peak
in the data, in the $\Sigma\pi$ decay channels two additional
Breit-Wigner functions for the $\Sigma(1385)$ and $\Lambda(1405)$
were necessary.

\begin{figure*}[ht!]
\hspace{0.5cm}
\resizebox{1.0\textwidth}{!}{%
\begin{tabular} {c c}
  \rotatebox{0}{\includegraphics{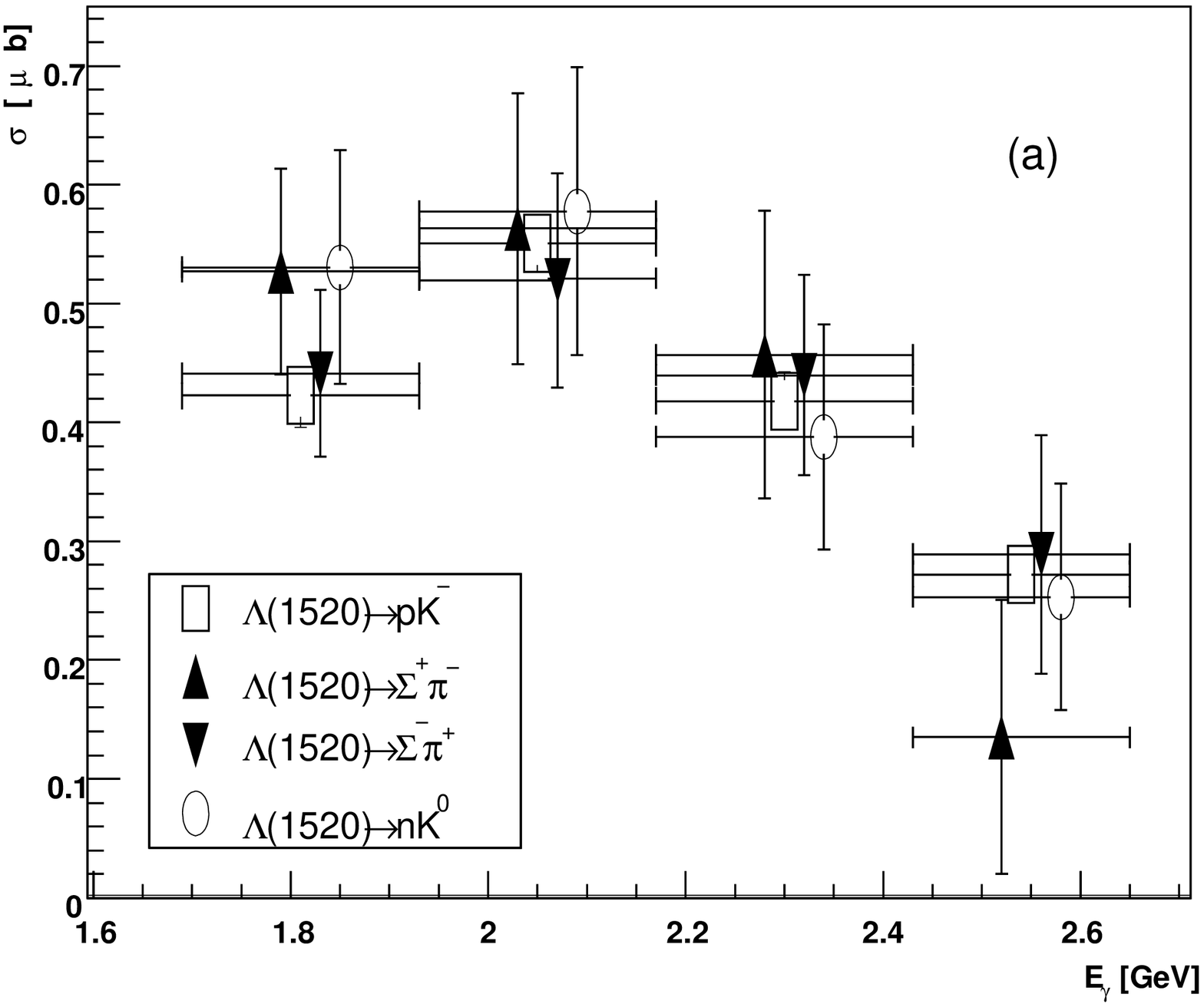}}\hspace{0.3cm}&
  \rotatebox{0}{\includegraphics{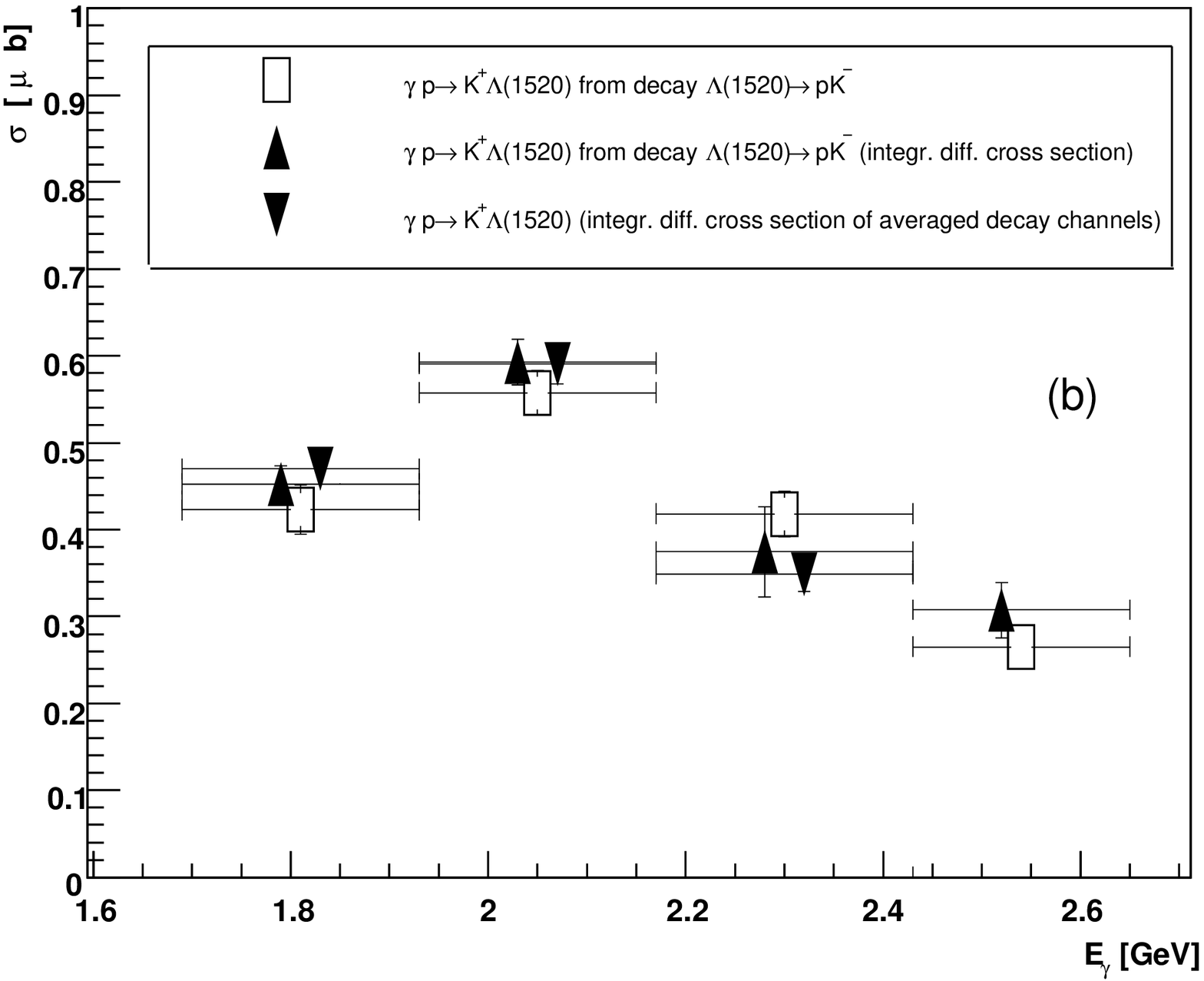}}\hspace{0.3cm}\\
\end{tabular}
} \caption{(a) Total cross section for the reaction $\gamma p
   \rightarrow K^+
  \Lambda(1520)$ as determined in different decay channels, (b) Comparison of
  the total cross sections for the dominant decay channel
  $\Lambda(1520)\rightarrow pK^-$ (see (a), squares) gained via integration of
  the differential cross sections d$\sigma$/d$t$ (upward triangles), and the
  averaged and integrated differential cross sections (downward triangles)
  from the four decay channels presented in (a).}
\label{pic:totcross}
\end{figure*}

The total cross sections $\sigma(E_{\gamma})$ for $\gamma p
\rightarrow K^+\Lambda(1520)$ determined from each decay channel was
calculated from the number
$N_{\Lambda(1520)}(E_{\gamma},\cos\theta_{K^+})$, the acceptance
$A(E_{\gamma},\cos\theta_{K^+})$, the branching ratio $BR$ and the
global luminosity as the product of the photon flux
$N_{\gamma}(E_{\gamma})$ through the target and the target area
density $\rho_T$:
\begin{center}
\begin{math}
\sigma(E_{\gamma})=\frac{1}{BR\cdot N_{\gamma}(E_{\gamma}) \rho_T}
\sum\frac{N_{\Lambda(1520)}(E_{\gamma},\cos\theta_{K^+})}{A(E_{\gamma},\cos\theta_{K^+})},
\end{math}
\end{center}
the sum extending over the $\cos\theta_{K^+}$ bins.

The results are shown in table \ref{tab:bg_beitraege_pkk}. The
quoted errors were calculated by adding in quadrature the
statistical errors, the systematic uncertainties due to the
parametrisation of the background underneath the $\Lambda(1520)$
peak, and the branching ratios. The total cross sections of the
reaction $\gamma p \rightarrow K^+\Lambda(1520)$ are shown in fig.
\ref{pic:totcross}. In fig. \ref{pic:totcross}(a) cross sections for
different decay channels are presented. In fig.
\ref{pic:totcross}(b) the results for the $\Lambda(1520)\rightarrow
K^-p$ decay channel determined by integration of $d\sigma/dt$ (see
section \ref{sec:diffcs}) is shown. The $d\sigma/dt$ cross sections
from the four decay channels were averaged and integrated. The
results are also shown in fig. \ref{pic:totcross}(b). The error bars
indicate the errors of the fitting procedure only, the differences
between the three methods indicate the magnitude of the systematic
errors. The results are consistent within the error bars.

\begin{table}[h!]
\caption{\label{tab:bg_beitraege_pkk}Total cross section for the
reaction $\gamma p \rightarrow K^+ \Lambda(1520)$ determined in
different decay channels of the $\Lambda(1520)$. The errors are
largely due to the systematical uncertainty of the background
parametrisation.}
\begin{center}
\renewcommand{\arraystretch}{1.2}
\begin{tabular}{cccc}
\hline\hline
$E_\gamma$ [GeV] & $pK^-$ [$\mu$b] & $\Sigma^+\pi^-$ [$\mu$b] & \\ \hline
1.69--1.93 & 0.422 $\pm$ 0.031 & 0.518 $\pm$ 0.084 & \\
1.93--2.17 & 0.503 $\pm$ 0.033 & 0.564 $\pm$ 0.113 & \\
2.17--2.41 & 0.418 $\pm$ 0.029 & 0.456 $\pm$ 0.118 & \\
2.41--2.65 & 0.270 $\pm$ 0.031 & 0.139 $\pm$ 0.115 & \\ \hline\hline
$E_\gamma$ [GeV] & $\Sigma^-\pi^+$ [$\mu$b] & $nK^0$ [$\mu$b] & average [$\mu$b]\\ \hline
1.69--1.93 & 0.442 $\pm$ 0.070 & 0.532 $\pm$ 0.097 & 0.442 $\pm$ 0.026 \\
1.93--2.17 & 0.519 $\pm$ 0.088 & 0.577 $\pm$ 0.118 & 0.482 $\pm$ 0.029 \\
2.17--2.41 & 0.439 $\pm$ 0.083 & 0.389 $\pm$ 0.098 & 0.420 $\pm$ 0.026 \\
2.41--2.65 & 0.297 $\pm$ 0.096 & 0.254 $\pm$ 0.088 & 0.242 $\pm$ 0.027 \\ \hline\hline
\end{tabular}
\renewcommand{\arraystretch}{1.0}\end{center}
\end{table}

\subsection{Differential cross sections}
\label{sec:diffcs}
\begin{figure*}[ht!]
\hspace{0.5cm}
\resizebox{0.85\textwidth}{!}{%
\begin{tabular} {c}
  \rotatebox{0}{\includegraphics{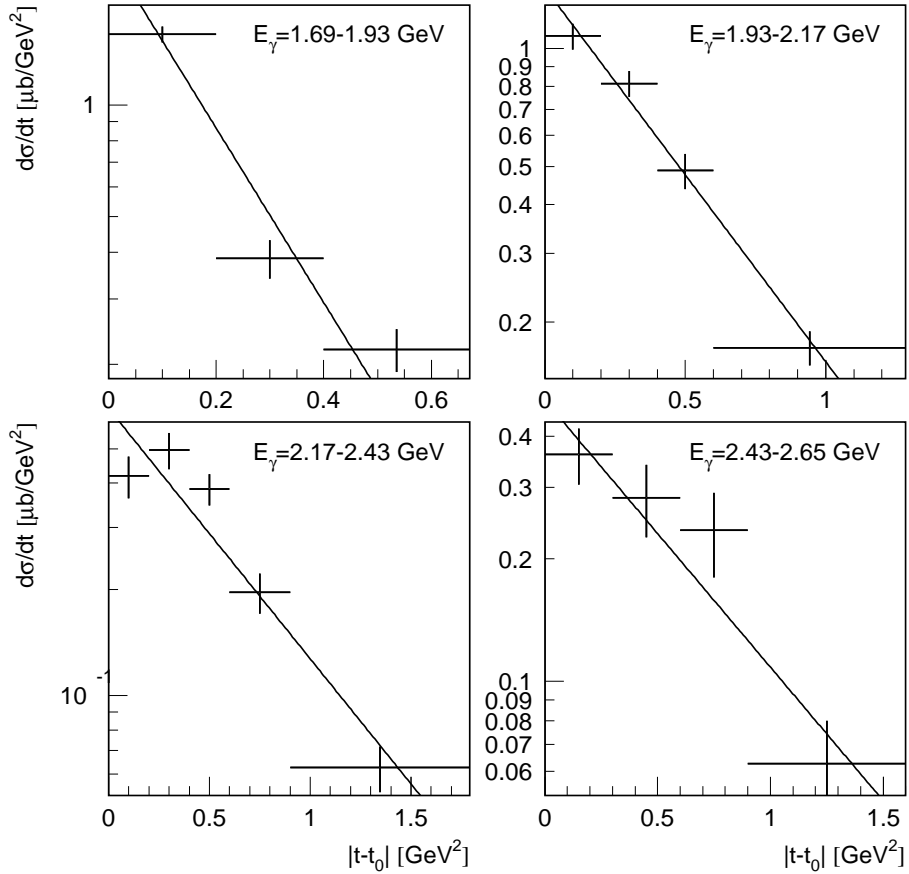}}
\end{tabular}
}
\caption{Differential cross section for the reaction $\gamma p \rightarrow K^+
  \Lambda(1520)$ determined via the decay channel $\Lambda(1520)\rightarrow
  pK^-$ in four photon energy bins as a function of $|t-t_0|$; $t_0$ denotes the minimal
  kinematically allowed squared four-momentum transfer, which was calculated on an
  event-by-event basis.}
\label{pic:diffcrosssec1}
\end{figure*}

The differential cross sections d$\sigma$/d$t$ of the reaction
$\gamma p \rightarrow K^+ \Lambda(1520)$ were determined for four
photon-energy intervals. Invariant mass distributions were plotted
in each energy and $t$ interval and fitted in order to determine the
number of produced $\Lambda(1520)$ hyperons. The cross sections are
shown in fig. \ref{pic:diffcrosssec1} and table \ref{tab:diffcross}.
In the four analyzed photon energy intervals the modulus of the
exponential $t$-slope  decreases from 5.4 to 1.5\,GeV$^{-2}$ with
increasing photon energies.

\begin{table}[h!]
\caption{Differential cross sections and $t$-slopes for the reaction $\gamma p
  \rightarrow K^+ \Lambda(1520)$ determined via the decay channel
  $\Lambda(1520)\rightarrow pK^-$ in four photon energy bins. For the slopes
  estimates on the systematical errors are given.}
\label{tab:diffcross}
\begin{center}
\renewcommand{\arraystretch}{1.2}
\begin{tabular}{cccc}
\hline\hline
$E_\gamma$ [GeV] & t [GeV$^2$] & d$\sigma$/d$t$ [$\mu$b/GeV$^2$] & slope [GeV$^{-2}$]\\ \hline
1.69--1.93 & 0.0--0.2   & 1.547 $\pm$ 0.077 & -5.41 $\pm$ 0.7 \\
           & 0.2--0.4   & 0.386 $\pm$ 0.046 & \\
           & 0.4--0.671 & 0.219 $\pm$ 0.029 & \\\hline
1.93--2.17 & 0.0--0.2   & 1.076 $\pm$ 0.081 & -2.20 $\pm$ 0.3\\
           & 0.2--0.4   & 0.814 $\pm$ 0.062 & \\
           & 0.4--0.6   & 0.489 $\pm$ 0.050 &\\
           & 0.6--1.285 & 0.172 $\pm$ 0.017 & \\\hline
2.16--2.41 & 0.0--0.2   & 0.419 $\pm$ 0.057 & -1.63 $\pm$ 0.4\\
           & 0.2--0.4   & 0.496 $\pm$ 0.058 &\\
           & 0.4--0.6   & 0.385 $\pm$ 0.039 &\\
           & 0.6--0.9   & 0.196 $\pm$ 0.025 &\\
           & 0.9--1.790 & 0.063 $\pm$ 0.092 &\\\hline
2.41--2.65 & 0.0--0.3   & 0.361 $\pm$ 0.057 & -1.51 $\pm$ 0.4\\
           & 0.3--0.6   & 0.283 $\pm$ 0.057 &\\
           & 0.6--0.9   & 0.235 $\pm$ 0.055 &\\
           & 0.9--1.60  & 0.063 $\pm$ 0.017 &\\ \hline\hline
\end{tabular}
\renewcommand{\arraystretch}{1.0}\end{center}
\end{table}

\begin{figure}[pb]
\centerline{\resizebox{0.45\textwidth}{!}{\includegraphics{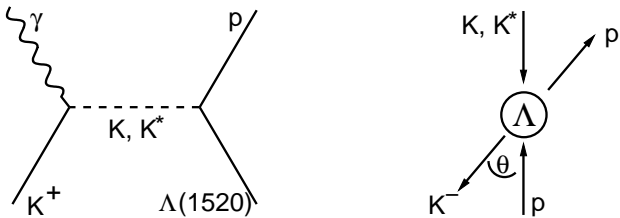}}\vspace{2mm}}
\caption{The $t$--channel helicity (Gottfried-Jackson) frame for the
$\Lambda(1520)$.\vspace{2mm} }
\label{pic:lambdavec}
\end{figure}

\begin{figure*}[ht!]
\hspace{0.5cm}
\resizebox{0.85\textwidth}{!}{%
\begin{tabular} {c}
  \rotatebox{0}{\includegraphics{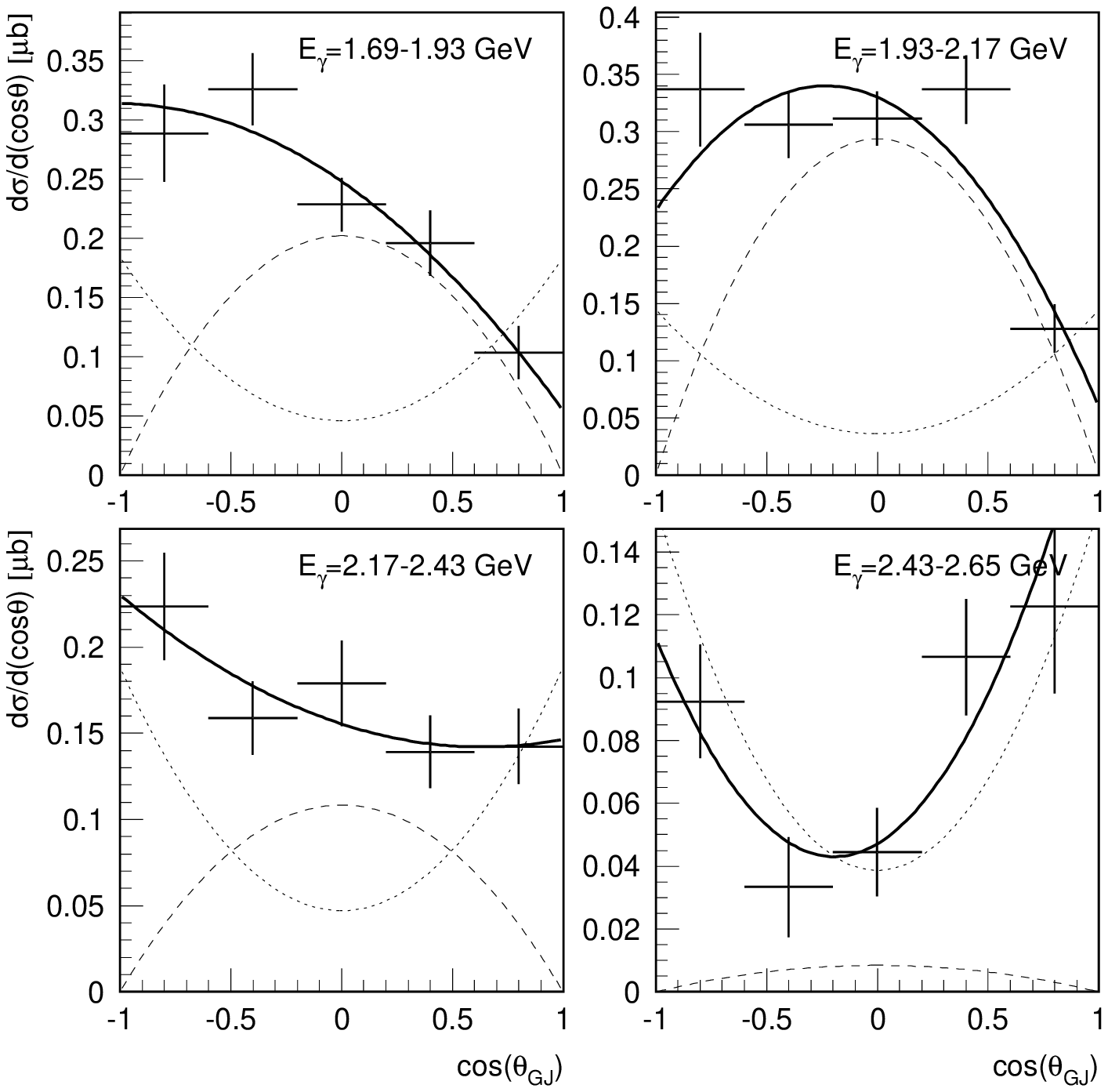}}\vspace{-3mm}
\end{tabular}
} \caption{The $\Lambda(1520)$ decay angular distribution in four
energy bins in the $t$--channel helicity system. Solid line:
complete fit (see text),
  dashed line: $1+3\cos^2\theta_{K^-}$ corresponding to $K^\star$ exchange,
  dotted line: $\sin^2\theta_{K^-}$  corresponding to $K$ exchange.}
\label{pic:diffcrosssec2}
\end{figure*}

The differential cross sections d$\sigma$/d$\cos\theta_\mathrm{GJ}$ in the $t$-channel helicity
frame (Gottfried-Jackson frame, fig.\,\ref{pic:lambdavec}) were determined from the decay
$\Lambda(1520)\rightarrow pK^-$, again in four photon energy bins
(fig. \ref{pic:diffcrosssec2}).


The $t$-channel helicity-frame $z$-axis is defined as antiparallel
to the incident photon in the $\Lambda(1520)$ rest frame, the
$y$-axis being normal to the production plane. The angular
distribution serves as an indicator in case of $t$-channel exchange.
If the $\Lambda(1520)$ ($J^P=3/2^-$) is produced with $m_z=\pm 1/2$
indicating $K^-$ exchange, one expects a $1+3\cos^2\theta_{K^-}$
distribution, whereas a $\sin^2\theta_{K^-}$ distribution indicates
$K^\star$ exchange with $m_z=\pm 3/2$. Published data from the LAMP2
Group \cite{lamp} strongly favour a nearly exclusive production via
$K^*$ exchange. The electroproduction data from CLAS
\cite{Barrow:2001ds} are presented in four $Q^2$ bins between 0.9
and 2.4\,GeV$^2$ averaged over $W$ from threshold to 2.43\,GeV. In
these data, $K$ exchange dominates, in particular towards larger
$Q^2$ values. The authors \cite{Barrow:2001ds} suggest that the
difference to the LAMP2 results might be due to substantial
contributions from longitudinally polarized photons. Our
photoproduction results between $E_\gamma$ = 1.69 and 2.65\,GeV do
not show a preference for $K^\star$ or  $K$ exchange in the
$t$--channel. We applied an $\alpha(1+3\cos^
2\theta_{K^-})+\beta\sin^2\theta_{K^-}+\gamma\cos\theta_{K^-}$ fit
in order to take into account interference terms from the $J=1/2$
hyperon production background, as suggested by the CLAS group
\cite{Barrow:2001ds}. When the data are divided into four energy
bins (fig. \ref{pic:diffcrosssec2}) a $\sin^2\theta_{K^-}$ dominance
is observed at low energies. For energies above 2.43\,GeV a
$1+3\cos^2\theta_{K^-}$ shape is seen which indicates that $K$
exchange is clearly preferred. This is in contrast to the LAMP2
results, but seems compatible with the CLAS electroproduction data.
The observed $t$-channel angular distribution changes rather rapidly
with energy. Hence it cannot be assigned to  variations in $K^\star$
or $K$ exchange. Therefore we conclude that this is likely due to
interference of $t$-channel exchange processes with amplitudes for
resonances formed in the $s$-channel.

\section{\boldmath $\Lambda(1520)$ decays\unboldmath}
\subsection{\boldmath Branching ratio for $\Lambda(1520)\to NK$ and
$\Sigma\pi$}
\begin{table}
\caption{\label{decays}$\Lambda(1520)$ decay branching ratios.}
\begin{center}
\renewcommand{\arraystretch}{1.3}
\begin{tabular}{lcc}
\hline\hline  Decay mode& PDG \cite{Amsler:2008zzb} & this work\\
\hline
$pK^-$ &\multirow{2}{*}{\} $45\pm1$\%}&  $(23.4\pm 1.6)\%$\\
$nK^0$ & & $(21.6\pm 2.8)\%$\\
$\Sigma^+\pi^-$ &\multirow{3}{*}{\} $42\pm1$\%}&  $(11.8\pm 2.1)\%$\\
$\Sigma^0\pi^0$ & & - \\
$\Sigma^-\pi^+$ & &  $(16.3\pm 1.7)\%$\\
\hline\hline
\end{tabular}\vspace{-6mm}
\renewcommand{\arraystretch}{1.0}\end{center}
\end{table}

The total and differential cross sections presented in tables
\ref{tab:bg_beitraege_pkk} and \ref{tab:diffcross} are calculated
using the $\Lambda(1520)$ decay branching fractions
\cite{Amsler:2008zzb}. In turn, the data allow us also to deduce the
branching fractions, even though with larger errors. The values
resulting from an integration of the efficiency corrected event
numbers are collected in table \ref{decays} and compared to PDG
numbers. The ratios presented here are normalized to the sum of the
fractions for decays into $pK^-\,+\,nK^0$ and
$3/2(\Sigma^-\pi^++\Sigma^+\pi^-)$, respectively. The branching
ratios are consistent with isospin invariance. The $\Sigma^-\pi^+$
and $\Sigma^+\pi^-$ branching ratios are compatible with 25\%
probability.

\subsection{The \boldmath $\Sigma(1385)\pi$ contribution to $\Lambda\pi\pi$
  in $\Lambda(1520)$ decays\unboldmath}

The $\Lambda\pi\pi$ decay channel is of special interest since the
$\Lambda(1520)$ might be a candidate for a $\Sigma(1385)\pi$
molecule. In \cite{Roca} it is argued that the $\Lambda(1520)$ has a
very strong coupling to the $\Sigma(1385)\pi$ channel, while the
small branching ratio into $\Lambda\pi\pi$ is assigned to the small
phase space of the decay into $\Sigma(1385)\pi$. With a strong
coupling to $\Sigma(1385)\pi$, the $\Lambda(1520)$ can be generated
dynamically in a unitarized coupled-channel calculation.

Experimentally, the fractional contribution of $\Sigma(1385)\pi$ in
$\Lambda(1520)\rightarrow\Lambda\pi\pi$ decays is not precisely
known. Watson et al. \cite{Watson:1963zz} found that this
contribution cannot be deduced reliably from their data. Mast et al.
\cite{Mast:1973cz} analyzed bubble chamber events of the type $K^-p
\rightarrow \Lambda\pi^+\pi^-$ and found that $0.82\pm 0.10$ of the
$\Lambda\pi^+\pi^-$ decay proceeds via $\Sigma(1385)\pi$. From
inelastic reactions of a $K^-$ beam impinging on a neutron (in a
D$_2$ target), Corden et al. \cite{Corden:1975zz} determined this
fraction to be $0.58 \pm 0.22$, while Burkhardt et al.
\cite{Burkhardt:1971tb} found $0.39 \pm 0.10$.

To determine this fractional contribution from the data presented
here, we define $\Sigma(1385)^{\pm}$ candidates by a 1.34 GeV $<
M_{\Lambda\pi^{\pm}} <$ 1.4\,GeV cut. For this purpose, the number
of $\Lambda(1520)$ hyperons in the missing mass distribution
recoiling against the $K^+$ is plotted in two ways (figs.
\ref{pic:lam1520Sigmaminuscut1} and
\ref{pic:lam1520Sigmaminuscut2}): first without any additional cuts
or anticuts, second with an anticut against the $\Sigma(1385)$
candidates. The number of observed $\Lambda(1520)$ hyperons is
nearly unchanged (see fig. \ref{pic:lam1520Sigmaminuscut1}). In fig.
\ref{pic:lam1520Sigmaminuscut2} the number of events without and
with a cut around the $\Sigma(1385)$ mass is shown. The number of
$\Lambda(1520)$ hyperons is largely reduced by this cut.  The
remaining events may be due to the decay chain
$\Lambda(1520)\to\Sigma(1385)^{\pm}\pi^{\mp}$ or due to some
background. We use the number of observed events as an upper limit
for the  ratio of $\Lambda(1520)$ partical decay widths
$\frac{\Gamma(\Sigma(1385)^{\pm}
\pi^{\mp}(\rightarrow\Lambda\pi^+\pi^-))}{\Gamma(\Lambda
\pi^+\pi^-)}$. We find this ratio to be less than 0.19 for a decay
via $\Sigma(1385)^- \pi^+$ and less than 0.25 for a decay via
$\Sigma(1385)^+ \pi^-$, yielding

\begin{center}
\begin{math}
\frac{\Gamma(\Sigma(1385)^{\pm} \pi^{\mp}(\rightarrow\Lambda\pi^+\pi^-))}{\Gamma(\Lambda
\pi^+\pi^-)} < 0.44 \quad {\rm (90\%\,c.\,l.)}
\end{math}
\end{center}

The result is hard to reconcile with the analysis of Mast et
al. \cite{Mast:1973cz} or the predictions of Roca et al. \cite{Roca}.

\section{Summary}
\label{sec:summary}

Photoproduction of the $\Lambda(1520)$ hyperon was studied from
threshold to 2.65\,GeV. The investigated decay channels were
$\Lambda(1520)  \rightarrow p K^-$, $\Lambda(1520) \rightarrow n
K^0_s$, $\Lambda(1520) \rightarrow K^+$ $\Sigma^+ \pi^-$,
$\Lambda(1520) \rightarrow K^+ \Sigma^- \pi^+$, and $\Lambda(1520)
\rightarrow \Lambda \pi^+ \pi^-$. We present total cross sections
for $\Lambda(1520)$ photoproduction obtained from the first four
decay channels. The four cross sections are fully compatible. The
differential cross section d$\sigma$/d$t$ for $\Lambda(1520)
\rightarrow pK^-$ is shown in four energy bins. For this channel
angular distributions in the $t$--channel helicity system
(Gottfried-Jackson system) were evaluated. Contrary to the LAMP2
\cite{lamp} results our measurements do not support dominance of
$K^*$ $t$--channel exchange. The results seem to be more compatible
with the CLAS \cite{Barrow:2001ds} electroproduction data. Therefore
we cannot support the conjecture of the CLAS collaboration that the
difference to the LAMP2 data might be due to the contribution of
longitudinally polarized photons. So, the deeper reason for the
discrepancy with the LAMP2 data has still to be understood, even
taking into account that the LAMP2 data were measured under
different kinematical conditions. The investigation of the
$\Lambda(1520) \rightarrow \Lambda \pi \pi$ channel shows no
indication for a dominant decay via $\Sigma(1385)\pi$.

\section{Acknowledgements}
\label{sec:acknoledgements}

We would like to thank the technical staff of the ELSA machine group
for their invaluable contributions to the experiment. We gratefully
acknowledge the support by the Deutsche Forschungsgemeinschaft in
the framework of the Schwerpunktprogramm ``Investigation of the
hadronic structure of nucleons and nuclei with electromagnetic
probes'' (SPP 1034 KL 980/2-3) and the Sonderforschungsbereich
SFB/TR16 (``Subnuclear Structure of Matter'').

%\end{sloppypar}


\end{document}